\documentclass[10pt,a4paper]{llncs}
\usepackage{graphicx}
\usepackage{subfigure}
\usepackage{amsmath}
\usepackage{amssymb}
\usepackage{multirow}
\usepackage{threeparttable}
\usepackage{hyperref}
\usepackage[a4paper]{geometry}
\usepackage{array}
\usepackage{url}
\def   \Mod   #1{\;{\rm mod}\;#1}

\newtheorem{claim_new}{Claim}
\renewcommand{\qed}{\hfill \ensuremath{\square}}
\newcommand{\qedTheorem}{\hfill \ensuremath{\blacksquare}}

\renewenvironment{proof}{
\vspace*{-\parskip}\noindent\textit{Proof.}}{$\qed$

\medskip
}

\newenvironment{proofTheorem}{
\vspace*{-\parskip}\noindent\textit{Proof.}}{$\qedTheorem$

\medskip
}

\begin{document}

\title{Secure Cloud Storage with Data Dynamics Using 
Secure Network Coding Techniques\thanks{A version of the paper has been published in IEEE Transactions on Cloud Computing
(available at \href{https://dx.doi.org/10.1109/TCC.2020.3000342}{IEEE Xplore}).
A preliminary version~\cite{Sengupta_ASIACCS} appeared in ASIACCS'16.
Part of work was done when B.~Sengupta and A.~Dixit were at Indian Statistical Institute, Kolkata, India.}}

\author{Binanda Sengupta\inst{1}
\and
Akanksha Dixit\inst{2}
\and
Sushmita Ruj\inst{3}}

\index{Sengupta, Binanda}
\index{Dixit, Akanksha}
\index{Ruj, Sushmita}
\institute{
Singapore Management University, Singapore\\
\email{binandas@smu.edu.sg}
\and
City, University of London, UK\\
\email{akanksha.dixit@city.ac.uk}
\and
CSIRO Data61, Australia and Indian Statistical Institute, Kolkata, India\\
\email{sushmita.ruj@data61.csiro.au}
}

\maketitle

\begin{abstract}
In the age of cloud computing, cloud users with limited storage can outsource their data to remote servers. 
These servers, in lieu of monetary benefits, offer retrievability of their clients' data at any point of time. 
Secure cloud storage protocols enable a client to check integrity of outsourced data. 
In this work, we explore the possibility of constructing a secure cloud storage for dynamic data 
by leveraging the algorithms involved in secure network coding. 
We show that some of the secure network coding schemes can be used to construct \textit{efficient} secure cloud storage protocols for dynamic data, 
and we construct such a protocol (DSCS I) based on a secure network coding protocol. 
To the best of our knowledge, DSCS I is the first secure cloud storage protocol for \textit{dynamic} data 
constructed using secure network coding techniques which is secure in the standard model. 
Although generic dynamic data support arbitrary insertions, deletions and modifications, 
\textit{append-only} data find numerous applications in the real world. 
We construct another secure cloud storage protocol (DSCS II) specific to append-only data --- 
that overcomes some limitations of DSCS I. 
Finally, we provide prototype implementations for DSCS I and DSCS II in order to evaluate their performance.

\end{abstract}

\section{Introduction}\label{sec:intro}

With the advent of cloud computing, cloud servers offer to their clients (cloud users)
various services that include delegation of huge amount of computation and
outsourcing large amount of data. 
For example, a client having a smart phone with a low-performance processor or 
limited storage cannot accomplish heavy computation 
or store large volume of data. 
Under such circumstances, she can delegate her computation/storage to the cloud server.

\begin{figure}[t]
\small
\centering
\fbox{\includegraphics[width=.48\textwidth]{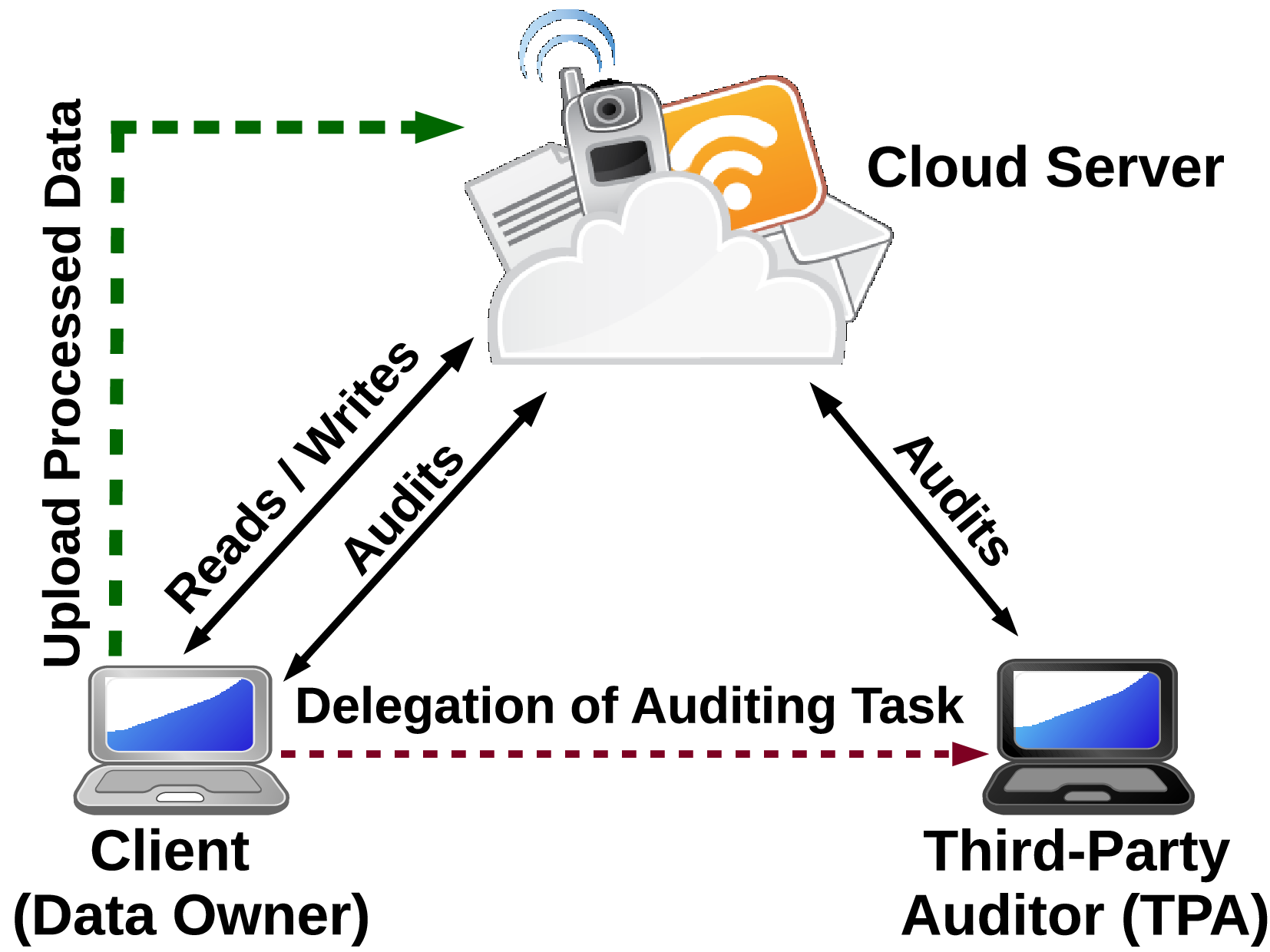}}
\caption{The entities involved in a secure cloud storage protocol.
	}\label{fig:dscs_j}
\end{figure}

In case of storage outsourcing, the cloud server stores  massive data on behalf
of its clients (data owners). However, a malicious cloud server can delete some of
the client's data (that are accessed infrequently) 
to save some space. Secure cloud
storage protocols (two-party protocols between the client and the server)
provide a mechanism to detect if the server stores the client's data untampered.
Based on the nature of the outsourced data, these protocols
are classified as: secure cloud storage protocols for \textit{static} data (SSCS)~\cite{Ateniese_CCS,JK_CCS,SW_JOC} 
and for \textit{dynamic} data (DSCS)~\cite{Erway_TISSEC,Wang_TPDS,Wichs_ORAM,Stefanov_CCS}.
For static data, the client cannot change her data
after the initial outsourcing (e.g., backup/archival data). 
Dynamic data are more generic in that the client can modify her data
as often as needed.
In secure cloud storage protocols, the client can \textit{audit} the outsourced data 
without accessing
the whole data file, and still be able to detect unwanted changes in data
done by a malicious server.
During an audit, the client sends a random challenge to the server which 
produces proofs of storage (computed on the stored data) corresponding to that challenge.
Secure cloud storage protocols are \textit{publicly verifiable} if an audit
can be performed by any third party auditor (TPA) using public parameters;
or \textit{privately verifiable} if an auditor needs some secret information of the client.
The entities involved in a secure cloud storage protocol and the interaction among them are shown in Figure~\ref{fig:dscs_j}.

In a network coding protocol~\cite{Ahlswede_IT,Li_IT},
each intermediate node (except sender/receiver nodes) on a network path
combines incoming packets to output another packet. 
These protocols
enjoy higher throughput, efficiency and scalability than 
the store-and-forward routing, 
but 
they are prone to \textit{pollution attacks} by 
malicious intermediate nodes injecting invalid packets. 
These 
packets produce more such packets downstream,
and the receiver might not 
finally decode the file sent by the sender node. 
Secure network coding (SNC) protocols
use cryptographic techniques to prevent these attacks:
the sender authenticates each packet 
by attaching a small tag to it. 
These authentication tags are generated using
homomorphic message authentication codes (MACs)~\cite{HMAC_ACNS}
or homomorphic signatures~\cite{Charles,BFKW_PKC,Gennaro_PKC,Dario_PKC}.
Due to homomorphic property, an 
intermediate node can combine incoming packets (and their tags)
into a packet and its tag. 

In this work, we look at the problem of constructing a secure cloud storage protocol for dynamic data (DSCS)
from a different perspective.
We investigate whether we can construct an efficient DSCS protocol using an SNC protocol.
In a previous work, Chen et al.~\cite{Sherman_IC} reveal a relationship between secure cloud storage and secure network coding.
In particular, they show that one can exploit some of the algorithms involved in an SNC protocol
in order to construct a secure cloud storage protocol for static data.
However, their construction does \textit{not} handle dynamic data --- that makes it insufficient in many applications
where a client needs to update (insert, delete or modify) the remote data efficiently.
Further investigations are needed towards an efficient DSCS construction 
using a secure network coding (SNC) protocol.

Network coding techniques have been used to construct distributed storage systems~\cite{Dimakis_IT,DD_POR}
where the client's data are disseminated across multiple servers. 
However, they primarily aim
to reduce the repair bandwidth when some of the servers fail.
On the other hand, we explore 
whether we can exploit the algorithms involved in an SNC protocol
to construct an efficient and secure cloud storage protocol for dynamic data (for a \textit{single} storage server).

Although dynamic data are generic in the sense that they support arbitrary update (insertion, deletion and modification) operations, 
append-only data (where new data corresponding to a data file are inserted only at the end of the file) find numerous applications as well. 
These applications primarily maintain archival as well as current data 
by appending the current data to the existing datasets. 
Examples of append-only data include data obtained from
CCTV cameras, ledgers containing monetary transactions, medical history of patients,
data stored at append-only databases, and so on. 
Append-only data are also useful for maintaining other
log structures (e.g., certificates are stored using append-only log structures
in certificate transparency schemes~\cite{laurie2013rfc}).
In many of such applications, the data owner requires a cloud server to store the bulk data 
in an untampered and retrievable fashion with append being the only permissible update.
Although secure cloud storage schemes for generic dynamic data also work for append-only data,
a more efficient solution (specific to append-only data files) would be helpful in this scenario.

\noindent{\bf Our Contribution}: \quad
Our major contributions in this work are summarized as follows.

\begin{itemize}
\item We 
explore the possibility
of providing a generic construction of a DSCS protocol from any SNC protocol.
We discuss the challenges for a generic construction in details and
identify some SNC protocols suitable for constructing efficient DSCS protocols.

\item 
We construct a publicly verifiable DSCS protocol (DSCS I) from an SNC protocol~\cite{Dario_PKC}. 
DSCS I handles dynamic data, i.e., a client can efficiently perform
updates (insertion, deletion and modification) on the outsourced data. 
We discuss the (asymptotic) performance and certain limitations of DSCS I.

\item 
We provide the formal security definition of a DSCS protocol and 
prove the security of DSCS I.

\item 
As append-only data are a special case of generic dynamic data, we can use DSCS I 
(which is based on~\cite{Dario_PKC}) for append-only data. 
However, we identify some SNC protocols that are \textit{not} suitable for building a secure cloud storage for generic dynamic data,
but efficient secure cloud storage protocols for append-only data can be constructed from them.
We construct such a publicly verifiable secure cloud storage protocol (DSCS II) for append-only data 
by using an SNC protocol proposed by Boneh et al.~\cite{BFKW_PKC}.

\item We discuss the (asymptotic) 
performance of DSCS II which overcomes some limitations of DSCS I.

\item We implement DSCS I and DSCS II and evaluate their performance based on storage overhead, computational cost and communication cost.
\end{itemize}

\section{Preliminaries and Background}
\label{prelims}
We denote an element $a$ chosen uniformly at random from a set $X$
by $a\xleftarrow{R}X$, a finite field by ${\mathbb{F}}$
and the security parameter by $\lambda$.
An algorithm $\mathcal{A}(1^\lambda)$ is probabilistic polynomial-time (PPT) 
if its running time is polynomial in $\lambda$ and its output $y$
depends on the internal coin tosses of $\mathcal{A}$.
A function $f:\mathbb{N}\rightarrow\mathbb{R}$ is negligible in $\lambda$ if $f(\lambda)<\frac{1}{\lambda^c}$,
for all positive integers $c$ and all sufficiently large $\lambda$.
For two integers $a$ and $b$ ($a\le b$), $[a,b]$ denotes the set $\{a,a+1,\ldots,b\}$.
The multiplication of a vector $\hbox{v}$ by a scalar $s$ is denoted by $s\cdot\hbox{v}$.

\subsection{Bilinear Maps}
\label{sec:blmap}
Let $G_1,G_2$ and $G_T$ be multiplicative cyclic groups of prime order $p$.
Let $g_1$ and $g_2$ be generators of $G_1$ and $G_2$, respectively.
A bilinear map (or pairing)~\cite{Galbraith_DAM} 
is a function $e: G_1\times G_2\rightarrow G_T$ such that:
\textbf{1}. for all $u\in G_1, v\in G_2, a,b\in{\mathbb{Z}}_p$, we have $e(u^a,v^b)=e(u,v)^{ab}$
(bilinear property);
\textbf{2}. $e$ is non-degenerate, i.e., $e(g_1,g_2)\not = 1$;
\textbf{3}. for all $u_1,u_2\in G_1, v\in G_2$, we have $e(u_1\cdot u_2,v)=e(u_1,v)\cdot e(u_2,v)$.

\subsection{Secure Network Coding}
\label{snc}
In a computer network, a sender (or source) node sends packets to a receiver (or target) node through intermediate nodes (or routers). 
In network coding~\cite{Ahlswede_IT}, 
each intermediate node 
encodes the incoming packets to form another packet and forwards this packet downstream.

\begin{figure}[t]
\centering
\fbox{\includegraphics[width=.48\textwidth]{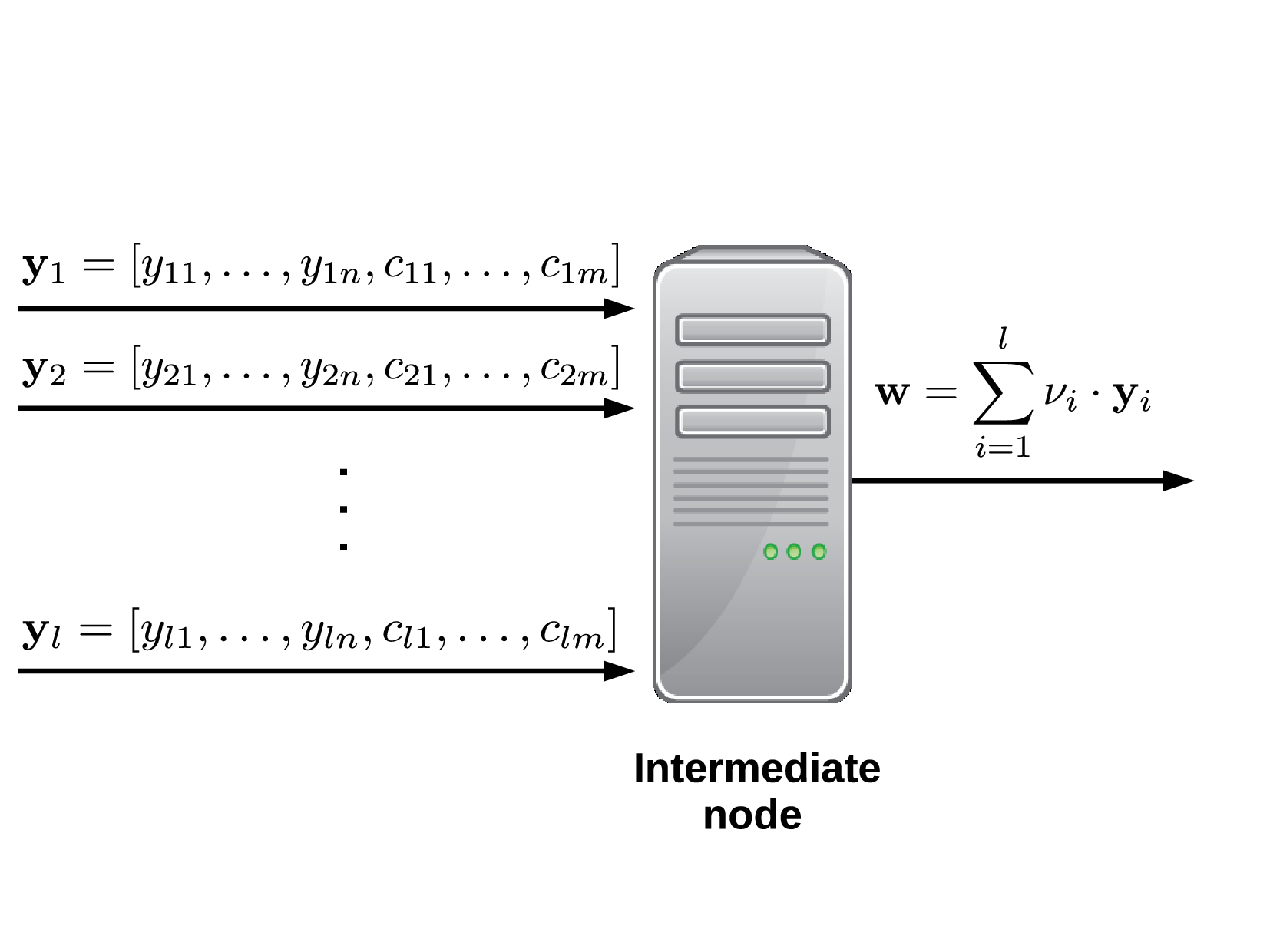}}
\caption{Linear combination of input packets for an intermediate node
	}\label{fig:snc}
\end{figure}

In this work, we consider \textit{random linear network coding}~\cite{RLNC1,RLNC2},
where each intermediate node encodes the incoming packets linearly using random coefficients.
We assume that each packet is a vector and each of its component is an element of a finite field ${\mathbb{F}}$.
Then, a file to be transmitted can be viewed as a set of $m$ vectors $\hbox{v}_1,\hbox{v}_2,\ldots,\hbox{v}_m\in{\mathbb{F}}^n$.
For each $i\in[1,m]$, the sender augments $\hbox{v}_i$ to form another vector
$\hbox{u}_i=[\hbox{v}_i~\hbox{e}_i]\in{\mathbb{F}}^{n+m}$, where $\hbox{e}_i$ is the $m$-dimensional
unit vector containing 1 in the $i$-th position and 0 in others,
and
transmits these augmented vectors (or packets).
Let $V\subset{\mathbb{F}}^{n+m}$ be the linear subspace spanned by 
$\hbox{u}_1,\hbox{u}_2,\ldots,\hbox{u}_m$.
A random file identifier \texttt{fid} is associated with the file (or $V$). 
Upon receiving $l$ packets $\hbox{y}_1,\hbox{y}_2,\ldots,\hbox{y}_l\in{\mathbb{F}}^{n+m}$,
an intermediate node 
chooses $l$ coefficients $\nu_1,\nu_2,\ldots,\nu_l\xleftarrow{R}{\mathbb{F}}$ and 
outputs a packet $\hbox{w}=\sum_{i=1}^{l}{\nu_i\cdot\hbox{y}_i}\in{\mathbb{F}}^{n+m}$ (as shown in Figure~\ref{fig:snc}).
This packet
is of the form
$\hbox{w}=[w_{1},w_{2},\ldots,w_{n},c_1,c_2,\ldots,c_m]\in V$,
where $c_1,c_2,\ldots,c_m\in{\mathbb{F}}$ and $w_j=\sum_{i=1}^{m}{c_i{v}_{ij}}$ for each $j\in[1,n]$.
Given $m$ linearly independent vectors, the receiver solves
a system of linear equations to obtain the original file. 

In a secure network coding (SNC) protocol, authentication tags are attached to packets,
which are 
computed using homomorphic MACs~\cite{HMAC_ACNS} 
or homomorphic signatures~\cite{Charles,BFKW_PKC,Gennaro_PKC,Dario_PKC}.
The \textit{homomorphic} property enables an intermediate node to combine the tags
of the incoming packets into a tag for the output packet.
We define an SNC protocol as follows~\cite{Dario_PKC}.
All algorithms
take
$m$ 
and
$n$ 
as inputs.
\begin{definition}[Secure Network Coding]\label{def:snc}
An SNC protocol consists of the following algorithms.
\begin{itemize}
 \item \emph{KeyGen$(1^\lambda,m,n)$:} The sender runs this algorithm in order to generate a secret key-public key pair
      $K=(sk,pk)$. 
 
 \item \emph{TagGen$(V,sk,m,n,{\texttt{fid}})$:} On input a linear subspace $V\subset{\mathbb{F}}^{n+m}$,
      the secret key $sk$ and a random file identifier \emph{\texttt{fid}} associated with $V$,
      the sender runs this algorithm to produce an authentication tag $t$ for $V$.
 
 \item \emph{Combine$(\{{\hbox{y}}_i,t_i,\nu_i\}_{1\le i\le l},pk,m,n,{\texttt{fid}})$:} Given
      $l$ incoming packets $\emph{\hbox{y}}_1,\emph{\hbox{y}}_2,\ldots,\emph{\hbox{y}}_l\in{\mathbb{F}}^{n+m}$ for 
      \emph{\texttt{fid}}
      and their corresponding tags $t_1,t_2,\ldots,t_l$,
      an intermediate node chooses $l$ coefficients $\nu_1,\nu_2,\ldots,\nu_l\xleftarrow{R}{\mathbb{F}}$
      and runs this algorithm. The algorithm outputs a
      packet $\emph{\hbox{w}}=\sum_{i=1}^{l}{\nu_i\cdot\emph{\hbox{y}}_i}\in{\mathbb{F}}^{n+m}$ and its tag $t$.
 
 \item \emph{Verify$({\hbox{w}},t,\bar{K},m,n,{\texttt{fid}})$:} Given a packet $\emph{\hbox{w}}$ for \emph{\texttt{fid}}
      and its tag $t$, an intermediate node or the receiver node
      executes this algorithm. This algorithm returns 1 if $t$ is a valid tag
      for the packet $\emph{\hbox{w}}$; it returns 0, otherwise.
 
\end{itemize}
\end{definition}

For some SNC protocols (e.g., \cite{BFKW_PKC,Gennaro_PKC,Dario_PKC}), verification requires
only the public key $pk$ (i.e., $\bar{K}=pk$).
For other SNC protocols (e.g., \cite{HMAC_ACNS}), the secret key $sk$ is required
for verification (i.e., $\bar{K}=sk$).

\medskip
\noindent
\textbf{Security of a Secure Network Coding Protocol}:\quad
The security of an SNC protocol based on homomorphic signatures is defined by 
the following security game
between a challenger and a probabilistic polynomial-time adversary $\mathcal{A}$~\cite{Dario_PKC}.
\begin{itemize}

\item \textbf{Setup}:\quad $\mathcal{A}$ provides the values $m$ and $n$ of its choice
to the challenger. The challenger runs KeyGen$(1^\lambda,m,n)$ to output $K=(sk,pk)$ and
returns $pk$ to $\mathcal{A}$.

\item \textbf{Queries}:\quad Let $q$ be a polynomial in $\lambda$.
The adversary $\mathcal{A}$ adaptively chooses a sequence 
of $q$ vector spaces $V_i\subset{\mathbb{F}}^{n+m}$ defined by respective augmented basis vectors
$\{\hbox{u}_{i1},\hbox{u}_{i2},\ldots,\hbox{u}_{im}\}$ and asks the challenger to authenticate
the vector spaces. For each $1\le i\le q$, the challenger chooses a random file identifier $\texttt{fid}_i$
from a predefined space, generates an authentication tag $t_i$ for $V_i$ by running TagGen$(V_i,sk,m,n,\texttt{fid}_i)$
and gives $t_i$ to $\mathcal{A}$.

\item \textbf{Forgery}:\quad The adversary $\mathcal{A}$ outputs $(\texttt{fid}^*,\hbox{w}^*,t^*)$.

\end{itemize}

Let the adversary $\mathcal{A}$ output the vector $\hbox{w}^*=[w^*_1,w^*_2,\ldots,w^*_{n+m}]\in {\mathbb{F}}^{n+m}$.
The adversary $\mathcal{A}$ wins the security game if:
$[w^*_{n+1},w^*_{n+2},\ldots,w^*_{n+m}]\in {\mathbb{F}}^m$ is not equal to the all-zero vector $0^m$,
Verify$(\hbox{w}^*,t^*,pk,m,n,\texttt{fid}^*)=1$ and
one of the following conditions is satisfied:\\
1. $\texttt{fid}^*\not=\texttt{fid}_i$ for all $i\in[1,q]$ (\textit{type-1 forgery})\\
2. $\texttt{fid}^*=\texttt{fid}_i$ for some $i\in[1,q]$, but $\hbox{w}^*\not\in V_i$ (\textit{type-2 forgery}).

For a secure network coding (SNC) protocol, the probability that the adversary $\mathcal{A}$ wins the security
game is negligible in the security parameter $\lambda$.

We note that the security game for a secure network coding protocol based on homomorphic MACs is same as
the game described above, except that the algorithm KeyGen now produces a secret key only
(unknown to $\mathcal{A}$) and the verification algorithm Verify requires
the knowledge of this secret key.

\subsection{Secure Cloud Storage}
\label{scs}

\noindent
\textbf{Provable Data Possession (PDP)}:\quad
Ateniese et al.~\cite{Ateniese_CCS} introduce \textit{provable data possession} (PDP)
where the client (data owner) splits the data file into blocks,
computes an authentication tag (e.g., MAC) for each block, 
and uploads the blocks along with their tags. 
During an audit, the client asks the server to prove the integrity of 
a predefined number of random blocks (\textit{challenge}).
The server computes a proof (\textit{response}) based on the challenge and the stored data, and sends
it to the client. 
A valid proof ensures retrievability of \textit{almost all} blocks in the file.
Ateniese et al.~\cite{Ateniese_CCS} also introduce the notion of public verifiability
where the client can delegate the auditing task to a third party auditor (TPA).
The TPA with the knowledge of the public key can perform an audit.
For privately verifiable schemes, only the client having 
the secret key can verify the proof sent by the server.
Other PDP schemes include~\cite{Ateniese_SCOM,Erway_TISSEC,Wang_TPDS,Wang_TC,FlexDPDP_EPR}.

\medskip
\noindent
\textbf{Proofs of Retrievability (POR)}:\quad
Juels and Kaliski~\cite{JK_CCS} introduce \textit{proofs of retrievability} (POR) for static data file
that ensures retrievability of \textit{all} of its blocks. 
According to Shacham and Waters~\cite{SW_JOC}, the
underlying idea is to encode the file
with an erasure code~\cite{MWSloane77,RSCode},
authenticate the blocks of the encoded file, and upload them on the server. 
With this technique, the server has to delete/modify a considerable
number of blocks to actually delete/modify a single block --- which can be detected with high probability.
Following the work by Juels and Kaliski, several POR schemes have been
proposed~\cite{Bowers_CCSW,Wichs_HA,IRIS,Wichs_ORAM,Stefanov_CCS,Bhavana_TCC,Outpor_CCS,Bowers_HAIL}.
Some of these schemes are designed for static data, and the rest allow the client
to modify data after the initial outsourcing.

We define a DSCS protocol 
as follows~\cite{Erway_TISSEC}.
It can be a PDP/POR protocol based on
the retrievability guarantee of data. 
The client (or a TPA) can be the \textit{verifier} (or auditor).

\begin{definition}[Secure Cloud Storage for Dynamic Data]\label{def:dscs}
A DSCS protocol consists of the following algorithms.

\begin{itemize}
\item \emph{KeyGen$(1^\lambda)$:} This algorithm generates a secret key-public key pair $K=(sk,pk)$
      for the client.

\item \emph{Outsource$(F,K,\texttt{fid})$:} The client splits the file $F$ associated with
      the file identifier \emph{\texttt{fid}} into $m$ blocks and
      computes authentication tags for them
      using her secret key $sk$. 
      She constructs an authenticated data structure $M$ on
      the tags (for checking freshness of the data) and computes some metadata $d_M$ for $M$.
      Finally, the client uploads $F'$ (the file $F$ and the tags)
      along with $M$ to the cloud storage server.
      She stores $d_M$ (and $m$) 
      or includes them in $pk$.
      
\item \emph{AuthRead$(i,F',M,pk,d_M,m,\texttt{fid})$:} When the client asks for the $i$-th block,
      the server sends her the $i$-th block $\emph{\hbox{v}}_i$, its 
      tag $t_i$
      and a proof $\Pi(i)$ (related to $M$) for $t_i$.

\item \emph{VerifyRead$(i,pk,d_M,m,\hbox{v}_i,t_i,\Pi(i),\texttt{fid})$:} 
      The client checks if $t_i$ is a valid tag for $\emph{\hbox{v}}_i$ and if $\Pi(i)$ matches with
      the latest metadata $d_M$. 
      She outputs 1 if both of them are satisfied; she outputs 0, otherwise.
      
\item \emph{InitUpdate$(i,{\texttt{updtype}},pk,d_M,m,{\texttt{fid}})$:} The variable
      \emph{\texttt{updtype}} indicates whether the update is an insertion (after) or a modification
      (of) or the deletion of the $i$-th block. 
      Depending on \emph{\texttt{updtype}},
      the client 
      asks the server to perform the required update
      on the file associated with \emph{\texttt{fid}} (update information is specified in \emph{\texttt{info}}).

\item \emph{PerformUpdate$(i,{\texttt{updtype}},F',M,{\texttt{info}},pk,{\texttt{fid}})$:}
      The cloud server performs the update on the file associated with \emph{\texttt{fid}} and sends the client
      a proof $\Pi$.

\item \emph{VerifyUpdate$(i,{\texttt{updtype}},{\texttt{info}},\Pi,pk,d_M,m,{\texttt{fid}})$:} Upon receiving the proof
      $\Pi$ for the file associated with \emph{\texttt{fid}}, 
      the client checks whether
      $\Pi$ is a valid proof.

\item \emph{Challenge$(pk,l,{\texttt{fid}})$:} During an audit, the verifier sends to the server
      a challenge set $Q$ of cardinality $l=O(\lambda)$ containing the block-indices she wants to audit.

\item \emph{Prove$(Q,pk,F',M,{\texttt{fid}})$:} The server 
      computes a proof of storage $T$ for $Q$ 
      and sends it to the verifier.

\item \emph{Verify$(Q,T,\bar{K},d_M,m,{\texttt{fid}})$:} The verifier checks if $T$ is a valid
      proof of storage for $Q$. 
      The verifier outputs 1 if both the proofs pass the verification; she outputs 0, otherwise.

\end{itemize}
\end{definition}
A DSCS protocol consists of the following sub-protocols:
\textsf{Init}, \textsf{Read}, \textsf{Write} and \textsf{Audit}.
The client performs \textsf{Init} that consists of the algorithms KeyGen and Outsource.
\textsf{Read} (that includes AuthRead and VerifyRead) and
\textsf{Write} (that comprises InitUpdate, PerformUpdate and VerifyUpdate)
are performed by the client and the server interactively.
The verifier interacts with the server via \textsf{Audit} that consists of Challenge,
Prove and Verify.
In a \textit{blockless verification}~\cite{Erway_TISSEC}, 
the proof $T$ includes a \textit{single aggregated block} corresponding to the blocks indexed by $Q$ ---
which reduces the communication bandwidth significantly.
For a privately verifiable DSCS protocol, $pk=\texttt{null}$ and $\bar{K}=sk$.
For a publicly verifiable DSCS protocol, $\bar{K}=pk$ and $pk$ includes both $m$ and $d_M$.

\subsection{Authenticated Data Structures Used in DSCS}
\label{skip_list}
Existing DSCS protocols use authenticated data structures 
(e.g., Merkle hash trees~\cite{Merkle_CR}, rank-based authenticated skip lists~\cite{Erway_TISSEC}
and rank-based RSA trees~\cite{Erway_TISSEC})
to ensure that the server stores
the latest version of the client's data.
The number of levels in a rank-based authenticated skip list~\cite{Erway_TISSEC}
is logarithmic in $m$ with high probability~\cite{Pugh_CACM}. 
Thus, the proof size, server computation time and client verification time are $O(\log m)$
with high probability.
We briefly discuss the algorithms of a rank-based authenticated skip list stored remotely in a server.
We refer to Appendix~\ref{app:skip_list} for details.
\begin{itemize}
\item
$\text{ListInit}(t_1,\ldots,t_m)$: Given an ordered list of elements $\{t_1,\ldots,t_m\}$,
the client constructs a rank-based authenticated skip list $M$
using a collision-resistant hash function $h$ to label its nodes.
She stores $m$ and the label of the root as the metadata $d_M$.

\item
$\text{ListAuthRead}(i,m,M)$: When the client asks the server for the $i$-th element $t_i$, 
the server sends the element along with a skip-list proof $\Pi(i)$ to the client.

\item
$\text{ListVerifyRead}(i,d_M,t_i,\Pi(i),m)$: Given 
$(t_i,\Pi(i))$, 
the client checks if the proof corresponds to the latest metadata $d_M$ stored at her end.
The client outputs 1 if the proof is valid; she outputs 0, otherwise.

\item
$\text{ListInitUpdate}(i,\texttt{updtype},d_M,t_i',m)$: The client initiates an update
to be performed on the skip list stored by the server.
An update can be an insertion after
or a modification of or the deletion of the $i$-th element. 
The type of the update is stored in a variable \texttt{updtype}. 
From the server's response, the client computes $d_M'$,
that should be the new metadata if the server performs the update correctly,
and updates $m$ (for insertion/deletion).
She stores $d_M'$ at her end temporarily
and asks
the server to perform the update specifying the location $i$, \texttt{updtype} 
and the new element $t_i'$ (\texttt{null} for deletion).

\item
$\text{ListPerformUpdate}(i,\texttt{updtype},t_i',M)$: Based on the value of \texttt{updtype},
the server performs the required update, 
computes a proof $\Pi$ similar
to the one generated during $\text{ListAuthRead}$ and sends $\Pi$ to the client.

\item
$\text{ListVerifyUpdate}(i,\texttt{updtype},t_i',d_M',\Pi,m)$: 
The client verifies $\Pi$ and computes $d_{new}$ from it. 
If $\Pi$ is a valid proof and $d_M'=d_{new}$, the client sets $d_M=d_M'$, 
deletes $d_M'$
and outputs 1. 
Otherwise, she changes $m$ to its previous value, 
deletes $d_M'$ and outputs 0.

\end{itemize}
Due to the collision-resistance property of $h$, 
the server cannot produce a valid proof $\Pi$ (during a read/update) without storing
the elements correctly.

\section{Construction of a DSCS Protocol Using an SNC Protocol}
\label{dscs}
Chen et al.~\cite{Sherman_IC} propose a generic construction of a secure cloud storage protocol
for \textit{static} data from any SNC protocol. 
They consider the data file $F$ 
to be a collection of $m$ blocks or vectors (each of dimension $n$). 
The client outsources the vectors
along with their authentication tags to the server.
Computing these tags exploits the algorithm SNC.TagGen.
During an audit, the client sends a subset $Q$ of $\{1,2,\ldots,m\}$ to the server.
The server augments the requested vectors, 
combines them linearly in an authenticated fashion using SNC.Combine (blockless verification) and
sends the output vector along with its tag to the client.
The client then verifies the authenticity of 
the received vector (using SNC.Verify).

In an SNC protocol, the number of vectors in the file to be transmitted is fixed,
as the dimension of the coefficient vectors used to augment the original vectors has to be known a priori. 
So, a \textit{generic} construction of a secure
cloud storage as described in~\cite{Sherman_IC} is suitable for static data only.
However,
clients in a DSCS protocol can update (insert, delete and modify) their data after the initial outsourcing.
We identify certain challenges towards providing a generic construction of an \textit{efficient} DSCS protocol
from an SNC protocol 
as follows.
We refer to Appendix~\ref{app:chal_dscs} for details.

\begin{enumerate}

\item The DSCS protocol must handle the varying values of $m$ appropriately
for insertions and deletions.

\item The index of a vector should not be embedded in its tag. 
Otherwise, for inserting or deleting a vector, 
the client has to recompute the tags for all subsequent vectors as their indices would change as well.

\item Freshness of the client's data must be guaranteed,
i.e., the server must store the up-to-date data.

\item Public verifiability is often desired where a third party auditor (TPA)
can audit on the client's behalf.

\end{enumerate}

\subsection{DSCS I: A DSCS Protocol Using an SNC Protocol}
\label{modified_scheme}

In this section, we construct a publicly verifiable DSCS protocol (DSCS I)
from the SNC protocol proposed by Catalano et al.~\cite{Dario_PKC}
which is secure in the standard model.
DSCS I uses a rank-based authenticated skip list 
to ensure the freshness of the dynamic data.
Let $h$ be the \textit{collision-resistant}
hash function used in the rank-based authenticated skip list. 
We assume that the file $F$ to be
outsourced is a collection of $m$ vectors (or blocks, according to Definition~\ref{def:dscs})
each of dimension $n$.
We note that a data block is the unit the file is split into, such that an authentication tag is assigned to each block
(thus, in this paper, a block represents a vector).
We call each of the $n$ components of a vector a segment.
We assume that each such segment is $\lambda$-bit long.
We describe the algorithms involved in DSCS I as follows.
We note that the algorithms KeyGen, Outsource, Prove and Verify in DSCS I call the algorithms
SNC.KeyGen, SNC.TagGen, SNC.Combine and SNC.Verify (respectively) of the underlying SNC protocol~\cite{Dario_PKC}
along with performing other operations related to the skip list. 

\begin{itemize}
\item
$\text{KeyGen}(1^\lambda,m,n)$: The client selects two random 
safe primes\footnote{A safe prime is of the form $2p'+1$, where $p'$ is also a prime.} $p,q$ 
such that 
$N=pq$ provides $\lambda$-bit security as an RSA modulus.
She takes a $(\lambda+1)$-bit random prime $e$
and the file identifier $\texttt{fid}=e$.
She selects $g,g_1,\ldots,g_n,h_1,\ldots,h_m\xleftarrow{R}{\mathbb{Z}}_N^*$.
The public key $pk$ is $(N,e,g,g_1,\ldots,g_n,h_1,\ldots,h_m,d_M,m,n)$.
The secret key $sk$ is $(p,q)$. 
Let $K=(sk,pk)$ and $d_M=\texttt{null}$.

\item
$\text{Outsource}(F,K,\texttt{fid})$: The file $F$ (associated with \texttt{fid})
consists of $m$ vectors each having $n$ segments
such that a segment is a bit-string of length $\lambda$ (an element of ${\mathbb{F}}_e$).
For each $1\le i\le m$, the $i$-th vector
$\hbox{v}_i$ is of the form $[v_{i1},\ldots,v_{in}]\in{{\mathbb{F}}}_e^n$. 
For each $\hbox{v}_i$,
the client selects $s_i\xleftarrow{R}{\mathbb{F}}_e$ and computes $x_i$ such that
\begin{equation}\label{netcod:eqn1}
x_i^e = g^{s_i}(\prod_{j=1}^{n}g_j^{v_{ij}})h_i\Mod N.
\end{equation}
Now, $t_i=(s_i,x_i)$ acts as an authentication tag for $\hbox{v}_i$.
The client constructs a rank-based authenticated skip list $M$ on the tags
$\{t_i\}_{1\le i\le m}$ and computes the metadata $d_M$ (the label of the root node of $M$).
She updates $d_M$ in the public key $pk$ and uploads the file
$F'=\{(\hbox{v}_i,t_i)\}_{1\le i\le m}$ along with $M$ to the server.

\item $\text{AuthRead}(i,F',M,pk,\texttt{fid})$: 
      The client and the server execute $\text{ListAuthRead}(i,m,M)$ for the skip list $M$. 
      The server sends the $i$-th vector ${\hbox{v}}_i$, its 
      tag $t_i$
      and a skip-list proof $\Pi(i)$ for $t_i$ to the client.

\item $\text{VerifyRead}(i,pk,\hbox{v}_i,t_i,\Pi(i),\texttt{fid})$: 
      The client, given ${\hbox{v}}_i=[v_{i1},\ldots,v_{in}]$,
      $t_i=(s_i,x_i)$ and $\Pi(i)$, 
      outputs 1 if
      $\text{ListVerifyRead}(i,d_M,t_i,\Pi(i),m)=1$ and 
      \begin{equation}\label{netcod:VerifyRead1}
	x_i^e= g^{s_i}(\prod_{j=1}^{n}g_j^{v_{ij}})h_i\Mod N.
      \end{equation}
      The client outputs 0, otherwise.

\item
$\text{InitUpdate}(i,\texttt{updtype},pk,\texttt{fid})$: The value of the variable \texttt{updtype} indicates if
the update is an insertion after or a modification of or the deletion of the $i$-th vector.
The client performs one of the following operations 
and temporarily stores $d_M'$ at her end.
\begin{enumerate}
 \item For insertion, the client selects
       $h'\xleftarrow{R}{\mathbb{Z}}_N^*$, 
       generates the new vector-tag pair $(\hbox{v}',t')$ and
       runs ListInitUpdate on 
       $(i,\texttt{updtype},d_M,t',m)$.
       She sends $(h',\hbox{v}')$ to the server.
       
 \item For modification, the client generates the new vector-tag pair $(\hbox{v}',t')$,
        runs ListInitUpdate on $(i,\texttt{updtype},d_M,t',m)$,
        sends $\hbox{v}'$ to the server.
       
 \item For deletion, the client runs ListInitUpdate
       on 
       $(i,\texttt{updtype},d_M,t',m)$, where $t'$ is \texttt{null}.
\end{enumerate}

\item
$\text{PerformUpdate}(i,\texttt{updtype},F',M,h',\hbox{v}',t',pk,\texttt{fid})$:
We assume that, for efficiency, the server keeps a copy of the ordered list $\mathcal{L}$ of $h_j$
values for $1\le j\le m$.
The server performs one of the following operations.
\begin{enumerate}
 \item For insertion, the server sets $m=m+1$, inserts $h'$ in the $(i+1)$-th position
       of $\mathcal{L}$ 
       and inserts $\hbox{v}'$ after the $i$-th vector.
       The server runs ListPerformUpdate on 
       $(i,\texttt{updtype},t',M)$.
       
 \item For modification ($h'$ is \texttt{null}), the cloud server modifies the $i$-th vector
       to $\hbox{v}'$ and runs ListPerformUpdate$(i,\texttt{updtype},t',M)$.
       
 \item For deletion ($h',\hbox{v}'$ and $t'$ are \texttt{null}), the server sets $m=m-1$,
       deletes 
       $h_i$ 
       from $\mathcal{L}$
       and runs
       ListPerformUpdate$(i,\texttt{updtype},\texttt{null},M)$.
\end{enumerate}

\item
$\text{VerifyUpdate}(i,\texttt{updtype},t',d_M',\Pi,pk,\texttt{fid})$:
Upon receiving the proof from the server, the client performs
$\text{ListVerifyUpdate}(i,\texttt{updtype},t',d_M',\Pi,m)$.
If its output 
is 1, the client outputs 1 and updates her public
key (the latest values of $m,d_M$ and $h_j$ for $j\in[1,m]$) accordingly. Otherwise, she outputs 0.

\item
$\text{Challenge}(pk,l,\texttt{fid})$: During an audit, the verifier selects $I$, a random $l$-element subset of $[1,m]$.
Then, she generates a challenge set $Q=\{(i,\nu_i)\}_{i\in I}$, where each $\nu_i\xleftarrow{R}{\mathbb{F}}_e$.
She sends $Q$ to the server.

\item
$\text{Prove}(Q,pk,F',M,\texttt{fid})$: 
The server computes $s=\sum_{i\in I}\nu_is_i\Mod e$ and $s'=(\sum_{i\in I}\nu_is_i-s)/e$. 
For each $i\in I$, the server
forms $\hbox{u}_i=[\hbox{v}_i~\hbox{e}_i]\in{\mathbb{F}}_e^{n+m}$ by augmenting $\hbox{v}_i$ with the unit
coefficient vector $\hbox{e}_i$. 
It computes
$\hbox{w}=\sum_{i\in I}\nu_i\cdot\hbox{u}_i\Mod e\in{\mathbb{F}}_e^{n+m}$,
$\hbox{w}'=(\sum_{i\in I}\nu_i\cdot\hbox{u}_i-\hbox{w})/e\in{\mathbb{F}}_e^{n+m}$ and
\begin{equation}\label{netcod:eqn2}
x=\frac{\prod_{i\in I}x_i^{\nu_i}}{g^{s'}\prod_{j=1}^{n}g_j^{w'_j}\prod_{j=1}^{m}h_j^{w'_{n+j}}}\Mod N.
\end{equation}
Let $\hbox{y}\in{\mathbb{F}}_e^n$ be the first $n$ entries of $\hbox{w}$ and $t=(s,x)$. The server sends $T=(T_1,T_2)$
as a proof of storage for 
$Q$, where $T_1=(\hbox{y},t)$ and
$T_2=\{(t_i,\Pi(i))\}_{i\in I}$.

\item
$\text{Verify}(Q,T,pk,\texttt{fid})$: Using $Q=\{(i,\nu_i)\}_{i\in I}$ and $T=(T_1,T_2)$ sent by the server,
the verifier constructs a vector $\hbox{w}=[w_1,\ldots,w_n,w_{n+1},\ldots,w_{n+m}]\in{\mathbb{F}}_e^{n+m}$, where
the first $n$ entries of $\hbox{w}$ are same as those of $\hbox{y}$ and the $(n+i)$-th entry is
$\nu_i$ if $i\in I$ (0 if $i\not\in I$).
For each $i\in I$, the verifier checks if $\Pi(i)$ is a valid proof for $t_i=(s_i,x_i)$ with respect to $d_M$.
She computes $\bar{s}=\sum_{i\in I}\nu_is_i\Mod e$ and verifies whether $\bar{s}\stackrel{?}=s$.
Finally, she checks if 
\begin{align}\label{netcod:eqn3}
x^e\stackrel{?}=g^{s}\prod_{j=1}^{n}g_j^{w_j}\prod_{j=1}^{m}h_j^{w_{n+j}}\Mod N.
\end{align}
The verifier outputs 0 if any verification fails; she outputs 1, otherwise.

\end{itemize}

\noindent
\textbf{Correctness of Eqn.~\ref{netcod:eqn3}}: 
For each $i\in I$, $\hbox{v}_i$ is augmented with $\hbox{e}_i$
to form $\hbox{u}_i=[\hbox{v}_i~\hbox{e}_i]\in{\mathbb{F}}_e^{n+m}$.
So, we can rewrite Eqn.~\ref{netcod:eqn1} as
\begin{equation*}
x_i^e  = g^{s_i}(\prod_{j=1}^{n}g_j^{v_{ij}})h_i\Mod N
       = g^{s_i}\prod_{j=1}^{n}g_j^{u_{ij}}\prod_{j=1}^{m}h_j^{u_{i(n+j)}}\Mod N.
\end{equation*}
For an honest server storing the challenged vectors correctly, 
\begin{equation*} 
\begin{split}
x^e 	& = \frac{\prod_{i\in I}{(x_i^e)}^{\nu_i}}{\left(g^{s'}\prod_{j=1}^{n}g_j^{w'_j}\prod_{j=1}^{m}h_j^{w'_{n+j}}\right)^e}\Mod N\\
	& = \frac{g^{\sum_{i\in I}{\nu_is_i}}\prod_{j=1}^{n}g_j^{\sum_{i\in I}{\nu_iu_{ij}}}\prod_{j=1}^{m}h_j^{\sum_{i\in I}{\nu_iu_{i(n+j)}}}}{\left(g^{s'}\prod_{j=1}^{n}g_j^{w'_j}\prod_{j=1}^{m}h_j^{w'_{n+j}}\right)^e}\Mod N\\
	& = g^{{\sum_{i\in I}{\nu_is_i}}-es'}\prod_{j=1}^{n}g_j^{{\sum_{i\in I}{\nu_iu_{ij}}}-ew'_j}\prod_{j=1}^{m}h_j^{{\sum_{i\in I}{\nu_iu_{i(n+j)}}}-ew'_{n+j}}\Mod N\\
	& = g^{s}\prod_{j=1}^{n}g_j^{w_j}\prod_{j=1}^{m}h_j^{w_{n+j}}\Mod N.
\end{split}
\end{equation*}

\section{Security of DSCS I}
\label{security_DSCSI}

A secure DSCS protocol has the following properties~\cite{Erway_TISSEC,Stefanov_CCS}.

\noindent
\textbf{Authenticity}: 
The server cannot
produce a valid proof of storage $T'$ (corresponding to $Q$) 
without storing
the challenged vectors and 
tags untampered, except with
a probability negligible in $\lambda$.

\noindent
\textbf{Freshness}: 
The server must store
the up-to-date 
data file $F$.

\noindent
\textbf{Retrievability}: 
Given a 
PPT 
adversary $\mathcal{A}$ that correctly responds to $Q$ 
with a 
non-negligible probability,
there exists a polynomial-time extractor 
$\mathcal{E}$ that can extract (at least) the challenged
vectors (except with a negligible probability) by challenging $\mathcal{A}$ for a polynomial
(in $\lambda$) number of times and verifying $\mathcal{A}$'s responses.

\subsection{Security Model}
\label{security_model}
DSCS I offers the guarantee of dynamic provable data possession (DPDP)~\cite{Erway_TISSEC}.
The untrusted server (acting as a PPT adversary $\mathcal{A}$) can be malicious
exhibiting Byzantine behavior and corrupt the client's data arbitrarily,
i.e., it can apply updates 
of its choice. 
The \textit{data possession game} between a challenger $\mathcal{C}$
and $\mathcal{A}$ is as follows.

\begin{itemize}
\item $\mathcal{C}$ runs KeyGen to generate $(sk,pk)$ and gives $pk$ to $\mathcal{A}$.
$\mathcal{A}$ selects a file $F$ associated with the identifier \texttt{fid} to store.
$\mathcal{C}$ processes $F$ to form another file $F'$ with the help of $sk$ and
returns $F'$ to $\mathcal{A}$.
$\mathcal{C}$ stores only some metadata to verify the future updates. 

\item $\mathcal{A}$ adaptively chooses a sequence of operations defined by
$\{\texttt{op}_i\}_{1\le i\le q_1}$ ($q_1$ is polynomial in
the security parameter $\lambda$), where $\texttt{op}_i$ is an authenticated read, an authenticated update (write) or an audit.
$\mathcal{C}$ executes these operations on the file stored by $\mathcal{A}$.
For an update operation defined by $(\texttt{updtype},\texttt{info})$,
$\mathcal{C}$ verifies the proof (sent by $\mathcal{A}$) by running $\text{VerifyUpdate}$
and updates her metadata if and only if the proof passes verification.
$\mathcal{A}$ is notified about the result of verification for each $\texttt{op}_i$.
$\mathcal{A}$ can corrupt the file in an arbitrary
way during the execution of these operations, i.e., it can update any part
of the file that need not be the same as those specified in $\{\texttt{op}_i\}_{1\le i\le q_1}$.

\item Let $F^*$ be the final state of the file after $q_1$ operations. 
$\mathcal{C}$ has the latest metadata for the file $F^*$. 
$\mathcal{C}$ challenges $\mathcal{A}$ with
a random challenge set $Q$, and $\mathcal{A}$ returns a proof $T=(T_1,T_2)$ to $\mathcal{C}$.
$\mathcal{A}$ wins the game if the proof passes verification.
$\mathcal{C}$ can challenge $\mathcal{A}$ for $q_2$ (polynomial in $\lambda$) 
times  
to extract (at least) the challenged vectors of $F^*$.
\end{itemize}

\begin{definition}[Security of a DSCS Protocol]\label{def:security_dpdp}
A DSCS protocol is secure if, for any probabilistic polynomial-time
adversary $\mathcal{A}$ who can win the data possession game mentioned above with some non-negligible
probability, there exists a polynomial-time extractor $\mathcal{E}$ that can extract
(at least) the challenged vectors 
by interacting (via challenge-response) with $\mathcal{A}$
polynomially many times.
\end{definition}

\subsection{Security Analysis of DSCS I}
\label{security_ana}
We state and prove Theorem~\ref{theorem_DSCSI} to analyze the security of DSCS I
which depends on the security of the underlying SNC protocol~\cite{Dario_PKC}. 
The SNC protocol~\cite{Dario_PKC} in turn derives its security from the
Strong RSA assumption~\cite{StrongRSA_ECR97}.

\begin{theorem}\label{theorem_DSCSI}
Given that the hash function used to construct the rank-based authenticated skip list is
collision-resistant and the underlying network coding protocol is secure,
the DSCS I protocol
is secure in the standard model
according to Definition~\ref{def:security_dpdp}.
\end{theorem}

\begin{proofTheorem}
We use the following claims to prove Theorem~\ref{theorem_DSCSI}.

\begin{claim_new}\label{claim_authenticity}
Given that the hash function used to construct the rank-based authenticated skip list is
collision-resistant and the underlying network coding scheme is secure,
the authenticity of the data file is guaranteed in DSCS I.
\end{claim_new}

\begin{proof}
Authenticity of data demands that the cloud server, without storing the challenged vectors and
their respective authentication tags appropriately, cannot produce a valid response $T'=(T_1',T_2')=((\hbox{y}',t'),T_2')$
for a challenge set $Q=\{(i,\nu_i)\}_{i\in I}$ during the data possession game
(and during the extraction phase).
The data file $F$ with a random (but unique) \texttt{fid} is identified by
the augmented vectors $\hbox{u}_i=[\hbox{v}_i~\hbox{e}_i]\in{\mathbb{F}}_e^{n+m}$ for $i\in[1,m]$.
Let $T=(T_1,T_2)=((\hbox{y},t),T_2)$ be the response computed honestly for the same challenge set $Q$;
thus, Verify$(Q,T,pk,\texttt{fid})=1$.
We consider 
two cases and prove that
the adversary can generate neither a valid $T_1'$ (Case I) nor a valid $T_2'$ (Case II).

\noindent
\textbf{Case I}:\quad
We show that if there exists a PPT adversary $\mathcal{A}$
that can break the authenticity of DSCS I, the security of the underlying SNC protocol is compromised.

If possible, we assume that, when challenged with $Q=\{(i,\nu_i)\}_{i\in I}$ during the data possession game or the extraction phase,
the adversary $\mathcal{A}$ produces a valid (but incorrect) response $T'=(T_1',T_2)=((\hbox{y}',t'),T_2)$
such that $\hbox{y}'\not=\hbox{y}$ (where $T=(T_1,T_2)=((\hbox{y},t),T_2)$ is the correct response).
As $T'$ is a valid response, Verify$(Q,T',pk,\texttt{fid})=1$.
Let $\hbox{w}=[w_1,\ldots,w_n,w_{n+1},\ldots,w_{n+m}]\in{\mathbb{F}}_e^{n+m}$ be a vector, where
the first $n$ entries of $\hbox{w}$ are same as those of $\hbox{y}$ and the $(n+i)$-th entry is
$\nu_i$ if $i\in I$ (0 if $i\not\in I$).
Let $\hbox{w}'=[w_1',\ldots,w_n',w_{n+1}',\ldots,w_{n+m}']\in{\mathbb{F}}_e^{n+m}$ be another vector,
where the first $n$ entries of $\hbox{w}'$ are same as those of $\hbox{y}'$ and the $(n+i)$-th entry is
$\nu_i$ if $i\in I$ (0 if $i\not\in I$). Clearly, $\hbox{w}\not=\hbox{w}'$ (as $\hbox{y}\not=\hbox{y}'$).
We observe that the algorithm Verify executes
the algorithm SNC.Verify (see Eqn.~\ref{netcod:eqn3} in Section~\ref{modified_scheme}).
As Verify$(Q,T',pk,\texttt{fid})$ outputs 1, it follows that SNC.Verify$(\hbox{w}',t',pk,m,n,\texttt{fid})=1$
(otherwise, the algorithm Verify would output 0).
We consider only the case where $\hbox{y}\not=\hbox{y}'$.
We do not take into account the case where $\hbox{y}=\hbox{y}'$ but $t\not=t'$,
since the tag for a vector $\hbox{y}$ is unique (in this case, SNC.Verify$(\hbox{y},t',pk,m,n,\texttt{fid})$ outputs 0
--- which implies that Verify$(Q,T',pk,\texttt{fid})$ also outputs 0).

We also note that, for a given challenge set $Q$, the set of indices $I$ and the corresponding coefficients $\nu_i$
(for $i\in I$) are randomly chosen by the challenger (data possession game) or by the extractor (extraction phase).
As the correct values of the basis vectors $\hbox{u}_1,\ldots,\hbox{u}_m$ for $F$
are unique at a given point of time, their linear combination using fixed coefficients
($i$-th coefficient is $\nu_i$ or 0 depending on whether $i\in I$ or $i\not\in I$) is also unique.
This \textit{unique} linear combination is $\hbox{w}$ ($\not=\hbox{w}'$).
As the last $m$ entries of $\hbox{w}'$ are same as those of $\hbox{w}$,
it follows that $\hbox{w}'\not\in\text{span}(\hbox{u}_1,\ldots,\hbox{u}_m)$.

To sum up, the pair $(\hbox{w}',t')$ thus constructed for the data file 
$F$ (identified by \texttt{fid})
satisfies the following conditions:
$[w_{n+1}',w_{n+2}',\ldots,w_{n+m}']\in {\mathbb{F}}_e^m$ is not equal to the all-zero vector $0^m$,
SNC.Verify$(\hbox{w}',t',pk,m,n,\texttt{fid})$ outputs 1 and
$\hbox{w}'\not\in\text{span}(\hbox{u}_1,\ldots,\hbox{u}_m)$.
This implies a \textit{type-2 forgery} on the secure network coding protocol~\cite{Dario_PKC} we use in DSCS I
(security of an SNC protocol is discussed in Section~\ref{snc}).
However, since this network coding protocol is secure in the standard model,
the adversary cannot produce such a response $T_1'=(\hbox{y}',t')$, except with some probability
negligible in the security parameter $\lambda$.

\noindent
\textbf{Case II}\quad
In DSCS I, the hash function
$h$ used to compute the labels of the nodes in the rank-based authenticated
skip list $M$ is collision-resistant. 
To produce a valid skip-list proof $T_2'\not=T_2$
with respect to the latest metadata $d_M$ (the label of the root node of the skip list), 
the adversary has to find a collision for the hash function $h$ in
some level of the rank-based authenticated skip list.
As $h$ is taken to be collision-resistant,
the adversary can forge a skip-list proof only with a probability negligible in $\lambda$.

This completes the proof of Claim~\ref{claim_authenticity}.
\end{proof}

\begin{claim_new}\label{claim_freshness}
Given that the hash function used to construct the rank-based authenticated skip list is
collision-resistant, the freshness of the data file is guaranteed in the DSCS I protocol.
\end{claim_new}

\begin{proof}
Freshness of the data file is maintained using the rank-based authenticated skip list
built over the authentication tags.
For each update request made by the adversary during the data possession game, the challenger runs the algorithm InitUpdate
(which in turn calls the algorithm ListInitUpdate) 
to compute the new metadata $d'_M$ from the skip-list proof $\Pi$ provided by the adversary.
After the adversary performs the update, it sends another proof $\Pi'$ corresponding to the updated skip list.
Then, the challenger runs the algorithm VerifyUpdate
(which in turn calls the algorithm ListVerifyUpdate)
to compute $d_{new}$ from $\Pi'$ and to check if $d_M'\stackrel{?}=d_{new}$.
The challenger updates the latest metadata $d_M=d_M'$ if and only if $d_M'=d_{new}$ and $\Pi'$ is a valid proof.
If the adversary is able to make the challenger output 1 during some execution of VerifyUpdate without storing
the latest authentication tags, then it must have found a collision for the hash function $h$ in
some level of the rank-based authenticated skip list.
However, as $h$ is taken to be collision-resistant,
this event (forging a skip-list proof) occurs with a negligible probability.

On the other hand, for each audit during the data possession game or the extraction phase,
freshness of authentication tags is guaranteed by checking the validity of the skip-list proof $T_2=\{(t_i,\Pi(i))\}_{i\in I}$
with respect to the latest metadata $d_M$ (using the algorithm Verify).
The output of Verify is 1 if and only if $T_2$ is a valid skip-list proof.
Again, if the adversary is able to forge a skip-list proof (with respect to $d_M$) without storing
the latest authentication tags, then it must have found a collision for the hash function $h$ in
some level of the rank-based authenticated skip list ---
which occurs only with a negligible probability.

Finally, although the correct value of the set of basis vectors $\hbox{u}_1,\ldots,\hbox{u}_m$ for $F$
is unique at a given point of time, it is changed for each update.
The malicious adversary might discard modifications of some of these vectors 
and keep an older version of them (along with up-to-date authentication tags).
Thus, when challenged for some of these vectors, the adversary provides a proof $T_1=(\hbox{y},t)=(\hbox{y},(s,x))$ 
which is correct but computed on older data.
We note that the algorithm Verify computes $\bar{s}=\sum_{i\in I}\nu_is_i\Mod e$
($s_i$ values are obtained from $T_2=\{(t_i,\Pi(i))\}_{i\in I}$)
and checks whether $\bar{s}$ is equal to $s$ ($s$ is a part of $T_1$).
However, as the coefficients $\nu_i$ in the challenge set $Q$ are randomly chosen by the challenger, the value of $s$
would be equal to $\bar{s}=\sum_{i\in I}\nu_is_i\Mod e$ only with probability $1/e$ which is again negligible in $\lambda$ 
(since $e=\Theta(2^{\lambda+1})$).

This completes the proof of Claim~\ref{claim_freshness}.
\end{proof}

We define a polynomial-time extractor algorithm $\mathcal{E}$ that can extract (at least)
the challenged vectors (except with negligible probability) by interacting with an adversary $\mathcal{A}$
that wins the data possession game mentioned above with some non-negligible probability.
As DSCS I satisfies the \textit{authenticity}
and \textit{freshness} properties mentioned above,
$\mathcal{A}$ cannot produce a proof $T=(T_1,T_2)$ for a given challenge set $Q=\{(i,\nu_i)\}_{i\in I}$ without storing
the challenged vectors and their corresponding tags properly, except with some negligible probability.
This means that if the output of the algorithm Verify is 1 during the extraction phase, the vector
$\hbox{y}$ in the proof is the linear combination of the original data vectors $\hbox{v}_i$ for $i\in I$
using coefficients $\{\nu_i\}_{i\in I}$.

Suppose the extractor $\mathcal{E}$ wants to extract
$l$ vectors indexed by $J$. It challenges $\mathcal{A}$ with $Q=\{(i,\nu_i)\}_{i\in J}$. If the proof
is valid (checked using Verify), $\mathcal{E}$ initializes a matrix
$M_\mathcal{E}$ as $[\nu_{1i}]_{i\in J}$, where $\nu_{1i}=\nu_{i}$ for each $i\in J$.
The extractor challenges $\mathcal{A}$ for the same $J$ but with different random coefficients.
If the algorithm Verify outputs 1 and the vector of coefficients is linearly independent to
the existing rows of $M_\mathcal{E}$, then $\mathcal{E}$ appends this vector to $M_\mathcal{E}$ as a row.
The extractor $\mathcal{E}$ runs this algorithm until the matrix $M_\mathcal{E}$ has $l$ linearly independent rows.
So, the final form of the full-rank matrix $M_\mathcal{E}$ is $[\nu_{ji}]_{j\in[1,l], i\in J}$.
Then, the challenged vectors can be extracted with the help of Gaussian elimination.

This completes the proof of Theorem~\ref{theorem_DSCSI}.
\end{proofTheorem}

\subsection{Probabilistic Guarantees of DSCS I}
\label{app:prob_guaranteesDSCSI}
If the cloud server corrupts a constant (say, $\beta$) fraction of vectors present in a data file,
then the server passes an audit with probability $p_{cheat}=(1-\beta)^l$, where $l$ is
the cardinality of the challenge set $Q$. The probability $p_{cheat}$ is very small
for large values of $l$.
Typically, $l$ is taken to be $O(\lambda)$ in order to make the probability
$p_{cheat}$ negligible in $\lambda$. Thus, the verifier detects a malicious
server corrupting $\beta$-fraction of the data file with probability $p_{detect}=1-p_{cheat}=1-(1-\beta)^l$,
and it guarantees the integrity of \textit{almost all} vectors of the file.

\begin{table*}[!t]
\footnotesize
\setlength{\tabcolsep}{2.1pt}
\centering
 \caption{Comparison among secure cloud storage protocols achieving PDP guarantees}{
 \begin{tabular}{|c|c|c|c|c|c|c|}
 \hline
 Secure cloud storage & Type of & Computation & Computation & Communication & Public & Security \\
 protocols & data & for verifier & for server & complexity & verifiability & model \\  
 \hline\hline
 PDP~\cite{Ateniese_CCS} & Static & $O(1)$ & $O(1)$ & $O(1)$ & Yes & RO$^\dag$ \\
 \hline
 Scalable PDP~\cite{Ateniese_SCOM} & Dynamic$^\ddag$ & $O(1)$ & $O(1)$ & $O(1)$ & No & RO  \\
 \hline
 DPDP I~\cite{Erway_TISSEC} & Dynamic & $O(\log\tilde{m})$ & $O(\log\tilde{m})$ & $O(\log\tilde{m})$ & Yes$^\S$ & Standard  \\
 \hline
 DPDP II~\cite{Erway_TISSEC} & Dynamic & $O(\log\tilde{m})$ & $O(\tilde{m}^\epsilon\log\tilde{m})^{\star}$ & $O(\log\tilde{m})$ & Yes$^\S$ & Standard  \\
 \hline
 Wang et al.~\cite{Wang_TPDS} & Dynamic & $O(\log\tilde{m})$ & $O(\log\tilde{m})$ & $O(\log\tilde{m})$ & Yes & RO  \\
 \hline
 Wang et al.~\cite{Wang_TC} & Dynamic & $O(\log\tilde{m})$ & $O(\log\tilde{m})$ & $O(\log\tilde{m})$ & Yes & RO  \\
 \hline
 FlexDPDP~\cite{FlexDPDP_EPR} & Dynamic & $O(\log\tilde{m})$ & $O(\log\tilde{m})$ & $O(\log\tilde{m})$ & Yes$^\S$ & Standard  \\
 \hline
 Chen et al.~\cite{Sherman_IC} & Static & $O(1)$ & $O(1)$ & $O(1)$ & Yes & Standard  \\
 \hline
 DSCS I (in this work) & Dynamic & $O(\log m)$ & $O(\log m)$ & $O(\log m)$ & Yes & Standard  \\
 \hline
 DSCS II (in this work) & Dynamic$^\P$ & $O(1)$ & $O(1)$ & $O(1)$ & Yes & RO  \\
 \hline
 \end{tabular}}
 \label{tab:comparison_PDP}
\begin{tablenotes}
\item [] For simplicity, we exclude the security parameter
$\lambda$ from complexity parameters (for an audit).
The value $\tilde{m}$ denotes the number of blocks the data 
file is split into (such that an authentication tag is associated
with each block). For example, $\tilde{m}=m$ in our DSCS (I and
II) protocols, where $m$ denotes the number of vectors.
	The term $O(\tilde{n})$ is added implicitly to each complexity parameter,
	where $\tilde{n}$ is the size of each block.
	For example, $\tilde{n}=n$ in DSCS I and II,
	where a vector having $n$ segments is considered as a block.
	For all protocols, storage at the verifier side is $O(1)$,
	and storage at the server side is $O(|F'|)$ where $F'$ is the outsourced file.
	If $l$ is the cardinality of the challenge set and the server corrupts $\beta$
	fraction of the file, the detection probability $p_{detect}=1-(1-\beta)^l$
	for all the protocols (except, in DPDP II, $p_{detect}=1-(1-\beta)^{\Omega(\log\tilde{m})}$).
\item [] $\dag$ RO denotes the random oracle model~\cite{BR_RO}.
\item [] $\ddag$ Scalable PDP supports deletion, modification and append operations;
	 insertion of a block in an arbitrary location 
	 is not supported.
\item [] $\S$ Although the authors do not claim public verifiability explicitly,
	 the protocol can be made publicly verifiable by 
	 making $d_{\tilde{M}}$ and $\tilde{m}$ public. 
\item [] $\star$ $\epsilon$ is a constant such that $0<\epsilon<1$.
\item [] $\P$ DSCS II supports only append operations; (arbitrary) insertion, deletion and modification 
	 operations are not supported.
\end{tablenotes}
\end{table*}

\section{Performance Analysis of DSCS I}
\label{perform_ana}

\subsection{Efficiency of DSCS I}
\label{efficiency}
The computational cost of 
DSCS I is dominated by the cost of exponentiations
(modulo $N$).
To generate $x$ in the tag for a vector, 
the client has to perform a multi-exponentiation~\cite{MultiEXP_SAC}
and calculate the $e$-th root of the result (see Eqn.~\ref{netcod:eqn1}). 
The server requires two multi-exponentiations to calculate $x$
(see Eqn.~\ref{netcod:eqn2}). 
To verify a proof using Verify, the verifier has to perform a multi-exponentiation
and a single exponentiation (see Eqn.~\ref{netcod:eqn3}).
Due to the properties of a skip list~\cite{Pugh_CACM},
the size of each proof $\Pi$ (related to the rank-based authenticated skip list), the time required
to generate $\Pi$ and the time required to verify $\Pi$ are $O(\log m)$ with high probability.
As DSCS I provides provable data possession (PDP) guarantees,
we compare DSCS I with other PDP schemes based on different parameters related to an audit as shown in Table~\ref{tab:comparison_PDP}.

\subsection{Limitations of DSCS I}
\label{limits_dscs}
We discuss a few limitations of DSCS I compared to DPDP I~\cite{Erway_TISSEC} (specifically)
as both of them handle \textit{dynamic} data, offer
\textit{public verifiability} and are secure in the \textit{standard model}.

\begin{enumerate}
\item The size of the public key in DSCS I is $O(m+n)$, whereas that 
in DPDP I is constant.

\item A tag in DSCS I is of the form $(s,x)$, where $s\in {\mathbb{F}}_e$
and $x\in{\mathbb{Z}}_N^*$. A tag in DPDP I is an element of ${\mathbb{Z}}_N^*$.
Thus, the size of a tag in DSCS I is larger than that in DPDP I by $\lambda+1$ bits
(as $e$ is a $(\lambda+1)$-bit prime).

\item In DSCS I, the values of $d_M$, $m$ and $h_i$ in the public key must be
changed for each insertion or deletion (a modification requires changing only the value of $d_M$),
whereas only the values of $d_{\tilde{M}}$ and $\tilde{m}$ need to be changed in DPDP I.
However, if the server keeps a local copy of the public key (an ordered list containing
$h_i$ values for $i\in [1,m]$), then small changes are required at the server side. The server inserts
the new $h$ value (sent by the client) in $(i+1)$-th position in the list (for insertion) or discards
the $i$-th $h$ value (for deletion).

\end{enumerate}

We note that the existing secure cloud storage protocol for static data~\cite{Sherman_IC} based on the same SNC protocol~\cite{Dario_PKC}
also suffers from the first two limitations mentioned above. 
However, in this paper,
we explore whether a DSCS protocol can be constructed from an SNC protocol. 
A more efficient (in terms of the size of the public key or 
a tag) SNC protocol can lead to 
a more efficient DSCS protocol. 
In the following section, we propose another DSCS protocol (DSCS II) for append-only data
that is much more efficient than DSCS I.

\section{More Efficient Solutions for Append-only Data}
\label{append}

In this section, we identify SNC protocols~\cite{BFKW_PKC,Gennaro_PKC} that can be used 
for constructing DSCS protocols for \textit{append-only data}, and we indeed provide an efficient DSCS protocol (DSCS II) for append-only data
using an SNC protocol proposed by Boneh et al.~\cite{BFKW_PKC}
(Appendix~\ref{snc_exmpl} discusses this SNC protocol).
As discussed in Appendix~\ref{app:chal_dscs},
this SNC protocol 
is \textit{not} suitable for constructing an efficient secure cloud storage for \textit{generic dynamic data}
as block-indices are embedded in tags.
However, this issue does not arise for append-only data where data blocks (or vectors) are 
inserted at the end only (thus the index of a new block does not affect that of an existing block).
Moreover, unlike~\cite{Dario_PKC}, the public-key size in~\cite{BFKW_PKC} does not depend 
on $m$, 
the number of vectors in the data file.
\textit{These two crucial observations lead us to construct a more efficient SNC-based DSCS protocol for append-only data}.

\subsection{DSCS II: A DSCS Protocol for Append-only Data}
\label{append_scheme}
We describe our DSCS II protocol for append-only data.
As append is the only operation allowed in DSCS II, 
we do not need \textit{freshness} of data (and authenticated data structures).
This is because existing data blocks are never updated for an append,
and there is no older but valid version of a data block that the server can retain.
For the same reason, DSCS II does not include VerifyUpdate
that was used in DSCS I. 
DSCS II consists of the following algorithms.

\begin{itemize}
\item
$\text{KeyGen}(1^\lambda,m,n)$: Let $\mathcal{G}=(G_1,G_2,G_T,e,\psi)$ be a bilinear group tuple,
where $G_1,G_2$ and $G_T$ are multiplicative cyclic groups of prime order $p>2^\lambda$, and the functions
$e: G_1\times G_2\rightarrow G_T$ (bilinear map) and $\psi:G_2\rightarrow G_1$ are efficiently
computable. The client selects $g_1,\ldots,g_n\xleftarrow{R} G_1\backslash\{1\}$,
$h\xleftarrow{R} G_2\backslash\{1\}$ and $\alpha\xleftarrow{R}{\mathbb{F}}_p$. She takes $z=h^\alpha$
and chooses a random file identifier \texttt{fid} $\in\{0,1\}^\lambda$.
Let $H: {\mathbb{Z}}\times{\mathbb{Z}}\rightarrow G_1$ be a hash function considered to be a random oracle~\cite{BR_RO}.
The public key is $pk=(\mathcal{G},H,g_1,\ldots,g_n,h,z,m,n)$, and the secret key is $sk=\alpha$.
Let $K=(sk,pk)$.

\item
$\text{Outsource}(F,K,\texttt{fid})$: The file $F$ (associated with
\texttt{fid}) consists of $m$ vectors each having $n$ segments.
Let each segment be an element of ${\mathbb{F}}_p$. 
Then, for each $1\le i\le m$, the client computes an authentication tag for the $i$-th vector
$\hbox{v}_i=[v_{i1},\ldots,v_{in}]\in{\mathbb{F}}_p^n$ as
\begin{align}\label{eqn_ap1}
t_i = \left(H(\texttt{fid}||i)\prod_{j=1}^{n}{g_j^{v_{ij}}}\right)^{\alpha}
\end{align}
(see Eqn.~\ref{eqn_Boneh} in Appendix~\ref{snc_exmpl}).
The client uploads the file
$F'=\{(\hbox{v}_i,t_i)\}_{1\le i\le m}$ to the cloud server.

\item $\text{AuthRead}(i,F',pk,\texttt{fid})$: In order to read the $i$-th vector,
      the client sends the index $i$ to the server.
      The server sends the $i$-th vector ${\hbox{v}}_i$ and its tag $t_i$.

\item $\text{VerifyRead}(i,pk,sk,\hbox{v}_i,t_i,\texttt{fid})$: Upon receiving ${\hbox{v}}_i=[v_{i1},\ldots,v_{in}]$ and $t_i$,
      the client outputs 1 if
      \begin{equation}\label{netcod:VerifyRead2}
	t_i = \left(H(\texttt{fid}||i)\prod_{j=1}^{n}{g_j^{v_{ij}}}\right)^{\alpha}.
      \end{equation}
      The client outputs 0, otherwise.

\item
$\text{InitUpdate}(pk,\texttt{fid})$: The client 
       sends a vector-tag pair $(\hbox{v}',t')$ to the server and sets $m=m+1$ in $pk$.

\item
$\text{PerformUpdate}(F',\hbox{v}',t',pk,\texttt{fid})$: The server inserts $\hbox{v}'$ after the $m$-th vector
       (i.e., at the end of the data file) and sets $m=m+1$.

\item
$\text{Challenge}(pk,l,\texttt{fid})$: During an audit, the verifier selects $I$, a random $l$-element subset of $[1,m]$.
She generates a challenge set $Q=\{(i,\nu_i)\}_{i\in I}$, where each $\nu_i\xleftarrow{R}{\mathbb{F}}_p$.
The verifier sends $Q$ to the server.

\item
$\text{Prove}(Q,pk,F',\texttt{fid})$:  
For each $i\in I$, the server 
forms $\hbox{u}_i=[\hbox{v}_i~\hbox{e}_i]\in{\mathbb{F}}_p^{n+m}$ by augmenting 
$\hbox{v}_i$ with the unit
coefficient vector $\hbox{e}_i$. It computes a tag
\begin{align}\label{eqn_ap2}
t=\prod_{i\in I}{t_i^{\nu_i}}
\end{align}
for the vector
$\hbox{w}=\sum_{i\in I}{\nu_i\cdot\hbox{u}_i}\in{\mathbb{F}}_p^{n+m}$.
Let $\hbox{y}\in{\mathbb{F}}_p^n$ be the first $n$ entries of $\hbox{w}$. 
The server sends $T=(\hbox{y},t)$ to the verifier
as a proof of storage with respect to $Q$.

\item
$\text{Verify}(Q,T,pk,\texttt{fid})$: Using $Q=\{(i,\nu_i)\}_{i\in I}$ and $T=(\hbox{y},t)$ sent by the server,
the verifier constructs a vector $\hbox{w}=[w_1,\ldots,w_n,w_{n+1},\ldots,w_{n+m}]\in{\mathbb{F}}_p^{n+m}$, where
the first $n$ entries of $\hbox{w}$ are same as those of $\hbox{y}$ and the $(n+i)$-th entry is
$\nu_i$ if $i\in I$ (0 if $i\not\in I$).
Finally, the verifier outputs 1 if
\begin{align}\label{eqn_ap3}
e(t,h)=e\left(\prod\limits_{j=1}^{m}{H(\texttt{fid}||j)^{w_{n+j}}}\prod\limits_{j=1}^{n}{g_j^{w_j}},z\right).
\end{align}
The verifier outputs 0, otherwise.

\end{itemize}
\noindent
\textbf{Correctness of Eqn.~\ref{eqn_ap3}}: 
For each $i\in I$, $\hbox{v}_i$ is augmented with $\hbox{e}_i$
to form $\hbox{u}_i=[\hbox{v}_i~\hbox{e}_i]\in{\mathbb{F}}_e^{n+m}$.
So, we can rewrite Eqn.~\ref{eqn_ap1} as
\begin{equation*}
t_i  = \left(H(\texttt{fid}||i)\prod_{j=1}^{n}{g_j^{v_{ij}}}\right)^{\alpha}\\
     = \left(\prod_{j=1}^{m}{H(\texttt{fid}||j)^{u_{i(n+j)}}}\prod_{j=1}^{n}{g_j^{u_{ij}}}\right)^{\alpha}.
\end{equation*}
\noindent
For an honest server storing the challenged vectors correctly, 
\begin{equation*}
\begin{split}
t 	= \prod_{i\in I}{t_i^{\nu_i}}
	& = \left(\prod_{j=1}^{m}H(\texttt{fid}||j)^{\sum_{i\in I}{\nu_iu_{i(n+j)}}}\prod_{j=1}^{n}{g_j^{\sum_{i\in I}{\nu_iu_{ij}}}}\right)^{\alpha}\\
	& = \left(\prod\limits_{j=1}^{m}{H(\texttt{fid}||j)^{w_{n+j}}}\prod\limits_{j=1}^{n}{g_j^{w_j}}\right)^{\alpha}.
\end{split}
\end{equation*}
Substituting the value of $t$ in $e(t,h)$, we get
\begin{equation*} 
\begin{split}
e(t,h) 	& = e\left(\left(\prod\limits_{j=1}^{m}{H(\texttt{fid}||j)^{w_{n+j}}}\prod\limits_{j=1}^{n}{g_j^{w_j}}\right)^{\alpha},h\right)\\
	& = e\left(\prod\limits_{j=1}^{m}{H(\texttt{fid}||j)^{w_{n+j}}}\prod\limits_{j=1}^{n}{g_j^{w_j}},h^{\alpha}\right)\\
	& = e\left(\prod\limits_{j=1}^{m}{H(\texttt{fid}||j)^{w_{n+j}}}\prod\limits_{j=1}^{n}{g_j^{w_j}},z\right).
\end{split}
\end{equation*}

\noindent
\textbf{Observations}:\quad
We make the following observations. 
\begin{itemize}

\item The DSCS II protocol supports only append operations at the end of the data file.

\item DSCS II is publicly verifiable in that
anyone with the knowledge of the public key can perform an audit.

\item As we have discussed earlier, since DSCS II supports only append operations on the data file,
the \textit{freshness} property is not required in DSCS II. 
Thus, DSCS II does not need authenticated data structures 
(e.g., rank-based authenticated skip lists)
in order to achieve freshness of data.
Therefore, we do not require the algorithm VerifyUpdate in DSCS II.

\item DSCS II is secure in the random oracle model~\cite{BR_RO}. 
The security game and the security proof of DSCS II are similar to that
of DSCS I, 
except that append is the only permissible update and 
that freshness is not required in DSCS II.
The guarantee of authenticity 
comes from the security of the underlying SNC protocol~\cite{BFKW_PKC}
that is secure in the random oracle model.

\end{itemize}

\subsection{Efficiency of DSCS II}
\label{effi_append}
While executing the algorithm Outsource, the client has to perform a multi-exponentiation
to generate the value of an authentication tag for each vector $t$
(see Eqn.~\ref{eqn_ap1}).
The server requires one multi-exponentiation to calculate the value of $t$
(see Eqn.~\ref{eqn_ap2} in the algorithm Prove).
The verifier has to perform two multi-exponentiations
and two pairing operations (see Eqn.~\ref{eqn_ap3})
to verify a proof using the algorithm Verify.
We note that each of the parameters ---
the size of a proof, the time required to generate a proof and the time required to verify a proof ---
is constant (independent of $m$) in DSCS II.
Different efficiency parameters of DSCS II related to an audit are shown in
Table~\ref{tab:comparison_PDP}. 
Detailed experimental evaluation of DSCS II is given in Section~\ref{sec:exp_DSCSII}.

An authentication tag $t$ in DSCS II belongs to $G_1$, and thus is
$2\lambda$ bits long~\cite{Galbraith_DAM}.
The DSCS II protocol for append-only data overcomes some of the limitations
of our DSCS I protocol described in Section~\ref{modified_scheme} as follows.
\begin{enumerate}
\item In DSCS II, the size of the public key is $O(n)$ (which is $O(m+n)$ in DSCS I),
where $n$ ($\ll m$) is decided during setup and kept unchanged during the protocol execution.

\item In DSCS II, only $m$ needs to be modified for each append operation ---
which is similar to DPDP I~\cite{Erway_TISSEC}. 

\end{enumerate}

\section{Experimental Results}
\label{sec:exp_results}

In this section, we evaluate the performance of DSCS I and DSCS II.

\subsection{Evaluation Methodology}
We measure
storage overhead (server), communication (between client and server) 
and computation cost (client and server) 
on a 2.5 GHz Intel i5 processor with 8GB RAM.
For cryptographic operations, 
we use the OpenSSL library 1.0.2~\cite{OpenSSL} for DSCS I
and the pairing-based cryptography (PBC) library 0.5.14~\cite{PBC} for DSCS II.
As
we use two different 
libraries to implement DSCS I and DSCS II,
the computation time taken by similar cryptographic operations (e.g., random number generation) varies for these protocols.

In our experiments, 
each 
file comprises $m$ data blocks (or vectors) each of size $n'$. 
We note that each data block in DSCS I and DSCS II comprises $n$ of data segments.
Thus, $n'=n\times s_{seg}$, where $s_{seg}$ is the size of each data segment.
In our experiments, we fix the value of $n'$ (block size) to be 500 KB,
and thus the value of $m$ (number of blocks) varies for files according to their size. 
We experiment with file sizes 1, 10, 50, 200 and 500 MB,
except in the comparison with~\cite{Sherman_IC}
given in Table~\ref{tab:comp_computation} and Table~\ref{tab:comp_storage}.
We take the security parameter $\lambda=112$ which is same as that considered in~\cite{Sherman_IC}.
The results reported in the following sections are taken to be the average by running the respective
experiments 50 times.
During an update (and an audit), the communication cost reported here includes only the size of tags and the size of proofs.
We note that these algorithms need to communicate a block (of constant size) also.
For ease of comparison, we do not take the block size into consideration (similar to~\cite{Sherman_IC}).

\subsection{Experimental Results for DSCS I}
\label{sec:exp_DSCSI}
We take $p$ and $q$ to be 1024-bit primes such that $N=pq$ is a 2048-bit RSA modulus that
provides 112-bit security.

\noindent
\textbf{Storage Overhead}:\quad
The client 
only needs to store her secret key and a metadata 
(the root-digest of the rank-based authenticated skip list) --- 
which incurs a constant storage cost. 
On the other hand, the server needs to store, along with the data file, all the authentication information 
including the skip list and the tags. 
This storage overhead 
is shown in Table~\ref{tab:storage_commn_dscsI}.  
We observe that, given a fixed block size $n'$, the percentage of additional storage remains almost constant 
as the file size increases. 
Thus, the server bears a trivial storage overhead even after 
supporting dynamic operations.

\begin{table}[t]
\setlength{\tabcolsep}{2.1pt}
\centering
\caption{Storage overhead for the server and communication cost in DSCS I}{
\begin{tabular}{|c|c|c|c|c|c|c|c|c|}
\hline
File size & Storage cost for tags & Storage & \multirow{2}{*}{$|Q|$} & \multicolumn{4}{c|}{Communication cost (KB)} \\
\cline{5-8}
(MB) & and skip list (KB) & overhead &   & Audit & Insert  & Modify & Delete  \\
\hline\hline
1 & 1.03 & 0.10\% & 2 & 3.58   & 2.54 & 2.28 & 2.00 \\
\hline
10 & 8.75  & 0.09\% & 10 & 16.83  & 2.54 & 2.28 & 2.00\\
\hline
50 & 39.25 & 0.08\% & 10 & 16.83  & 2.54 & 2.28 & 2.00\\
\hline
200 & 153.86 & 0.08\% & 112 & 185.74 & 2.54 & 2.28 & 2.00\\
\hline
500 & 382.34 & 0.08\% & 112 & 185.74 & 2.54 & 2.28 & 2.00\\
\hline
\end{tabular}}
\label{tab:storage_commn_dscsI}
\end{table}

\begin{table}[t]
\setlength{\tabcolsep}{1.20pt}
\centering
\caption{Computation cost for the client $C$ and the server $S$ in DSCS I}
\begin{tabular}{|c|c|c|c|c|c|c|c|c|c|c|c|}
\hline
{File } & {Outsource} & \multirow{3}{*}{$|Q|$} & {Challenge} & {Prove} & {Verify} &
\multicolumn{2}{c|}{Insert }&
\multicolumn{2}{c|}{Delete }&
\multicolumn{2}{c|}{Modify }\\
{size} & {(sec)} & & {(msec)} & {(sec)} & {(sec)} &
\multicolumn{2}{c|}{(sec)}&
\multicolumn{2}{c|}{(sec)}&
\multicolumn{2}{c|}{(sec)}\\
\cline{2-2}
\cline{4-12}
(MB) & $C$ &   & $C$ & $S$ & $C$ & $C$ & $S$ & $C$ & $S$ & $C$ & $S$  \\
\hline
\hline
1 & 10.03 & 2 & 0.01 & 9.52 & 5.15 & 5.50 & 0.01 & 0.001 & 0.001 & 5.41 & 0.01 \\
\hline
10 & 100.08 & 10 & 0.02 & 9.58 & 5.00 & 5.49 & 0.10 & 0.001 & 0.009 & 5.39 & 0.11 \\
\hline
50 & 501.30 & 10 & 0.02 & 9.76 & 5.07  & 5.46 & 0.49 & 0.002 & 0.489 & 5.45 & 0.49 \\
\hline
200 & 2008.93 & 112 & 0.23 & 14.63 & 5.08 & 5.50 & 6.29 &  0.002 & 6.287 & 5.50  & 6.28 \\
\hline
500 & 5717.74 & 112 & 0.29 & 15.61 & 5.07 & 5.52 & 13.04 &  0.002 & 13.016 & 5.45 & 13.00  \\
\hline
\end{tabular}
\label{tab:comp_dscsI}
\end{table}

\noindent
\textbf{Communication Cost}:\quad
During an audit, the communication cost 
depends on the number of challenged blocks (i.e., $|Q|$)
which is a small constant. 
In response to $Q$, 
the server sends proofs to the client. 
The proof size 
depends on two factors: the size of a constant aggregated block (along with that of an aggregated tag) and 
the size of the skip-list proofs for the queried blocks.
The second factor brings some variation in the proof size as the number of queried blocks changes. 
As we later compare the performance of DSCS I with that of~\cite{Sherman_IC}, 
we do not include the block size in the 
cost (similar to~\cite{Sherman_IC}). 
From Table~\ref{tab:storage_commn_dscsI}, we observe that, in spite of having data dynamics, DSCS I consumes low 
bandwidth.

During an update, the communication cost depends on the type of update. 
For an insertion, the client sends to the server an index, a public parameter $h$, the new data block along with its tag. 
For a modification, client only needs to send an index, the modified block and its tag. 
For a deletion, communication includes the index of the block to be deleted. 
For each update, the server returns a proof to the client.  
The communication cost is reported in Table~\ref{tab:storage_commn_dscsI} for each update.

\noindent
\textbf{Computation Cost}:\quad
We report the computation cost for 
the following phases of DSCS I:
outsourcing 
(client), 
challenge generation (client), 
proof generation (server), 
proof verification (client),
and updates on the outsourced file (client and server). 
The time for  outsourcing includes splitting the file into blocks, 
tag computation 
and building a skip list.

The experimental results have been shown in Table~\ref{tab:comp_dscsI}. 
It is evident that the initial outsourcing of the data file is computationally expensive, but it 
is a one-time process. It grows as the file size increases. 
It depends on several factors like the block size $n'$, the number of blocks $m$ in a file and the number of segments $n$ in each block. 
Given a data file and the size of each segment in a block, if $n'$ is taken to be large, 
the computation time for generating a single tag increases as there are more segments per block (i.e., more components in each vector). 
On the other hand, if $n'$ is taken to be small, the time taken for computing all the tags (and building the skip list) increases 
as the number of blocks $m$ increases and so does the number of tags.  
Therefore, an appropriate value results in a good trade-off between them.

The time for challenge generation is small as much computation is not involved in this phase. 
Proof generation time does not depend on the file size but on the number of blocks queried ($|Q|$).
This includes both the time for generating an aggregated block (along with the aggregated tag) and 
the time for computing the skip-list proof for each challenged block. 
The time for proof verification includes the time for matching the aggregated block with the aggregated tag and 
the time for verifying the skip-list proof for each challenged block. 
The computation time for updates is shown in Table~\ref{tab:comp_dscsI} for the client $C$ and the server $S$ separately.

\begin{table}[t]
\setlength{\tabcolsep}{1.2pt}
\centering
\caption{Comparison based on computation cost}
\begin{tabular}{|c|c|c|c|c|c|c|c|c|c|c|}
\hline
{File } & \multirow{3}{*}{$n'$} & \multirow{3}{*}{$m$} & 
\multicolumn{2}{c|}{Outsource} & 
\multicolumn{2}{c|}{Challenge} &
\multicolumn{2}{c|}{Prove} &
\multicolumn{2}{c|}{Verify}\\
{size}  & & & 
\multicolumn{2}{c|}{(sec)} & 
\multicolumn{2}{c|}{(msec)} &
\multicolumn{2}{c|}{(sec)} &
\multicolumn{2}{c|}{(sec)}\\
\cline{4-11}
(MB) & & & ~\cite{Sherman_IC} & DSCS I & ~\cite{Sherman_IC} & DSCS I & ~\cite{Sherman_IC} & DSCS I & ~\cite{Sherman_IC} & DSCS I \\
\hline
\hline
1.45  & 1 KB & 1488  & 1119.60 & 30.78 & 0.04 & 0.24 & 0.93 & 0.08&  0.73 & 0.04\\
\hline
23.5  & 1 KB & 24089 & 18409.61 & 1144.91 & 2.00 & 0.25 & 1.01 & 0.11 & 0.77 & 0.04\\
\hline
121   & 1 MB & 122   & 87137.90 & 1222.60 & 0.38 & 0.23 & 970.69 & 29.29 & 738.41 & 10.34\\
\hline
432   & 1 MB & 433   & 325039.36 & 4463.70 & 4.20 & 0.22 & 976.57 & 30.19 & 741.24 & 10.62\\
\hline
\end{tabular}
\label{tab:comp_computation}
\end{table}

\begin{table}[t]
\setlength{\tabcolsep}{2.1pt}
\centering
\caption{Comparison based on storage overhead and communication cost}
\begin{tabular}{|c|c|c|c|c|c|c|}
\hline
{File size} & \multirow{2}{*}{$n'$} & \multirow{2}{*}{$m$} & 
\multicolumn{2}{c|}{Storage overhead} & 
\multicolumn{2}{c|}{Communication cost (audit)} \\
\cline{4-7}
(MB) & & & ~\cite{Sherman_IC} & DSCS I & ~\cite{Sherman_IC} & DSCS I \\
\hline
\hline
1.45  & 1 KB & 1488  & 0.54 MB & 0.58 MB & 376 B & 185.74 KB  \\
\hline
23.5  & 1 KB & 24089 & 8.68 MB & 8.86 MB & 376 B & 185.74 KB \\
\hline
121   & 1 MB & 122   & 0.04 MB & 0.05 MB & 376 B & 185.74 KB \\
\hline
432   & 1 MB & 433   & 0.16 MB & 0.17 MB & 376 B & 185.74 KB\\
\hline
\end{tabular}
\label{tab:comp_storage}
\end{table}

\noindent
\textbf{Comparison between DSCS I and Existing SNC-based Secure Cloud Storage}:\quad
Chen et al.~\cite{Sherman_IC} use the SNC protocol proposed by Catalano et al.~\cite{Dario_PKC}
to construct a secure cloud storage for \textit{static} data.
Our DSCS I construction for \textit{dynamic} data also exploits the same SNC protocol.
We prepare two comprehensive comparison tables (Table~\ref{tab:comp_computation} and Table~\ref{tab:comp_storage}) 
by taking results from the work 
of Chen et al.~\cite{Sherman_IC} (reported for a 3.1 GHz Intel i3 processor with 4GB RAM) and the results obtained from our experiments.
The experiments are done for the same values of parameters (the file size and the values of $n'$ and $m$) as reported in~\cite{Sherman_IC}.
Table~\ref{tab:comp_computation} shows that 
the computation cost for DSCS I is much less than
that for~\cite{Sherman_IC}. 
This is possibly due to the difference in the architectures these two schemes are implemented on.
Here, we have considered only the cost for initial outsourcing and an audit
(as~\cite{Sherman_IC} does not support updates on the data file).

From Table~\ref{tab:comp_storage}, we observe that, compared to~\cite{Sherman_IC}, DSCS I demands
some extra storage that is attributed to the skip list (for handling the dynamic data efficiently). 
This results in a slight increase in the storage overhead at the server side. 
During an audit,
\cite{Sherman_IC} requires a (constant) communication cost of 376 B for different file sizes 
(as the proof consists of an aggregated block and its tag).
On the other hand, the proof sent by the server in DSCS I includes not only an aggregated block (and its tag) but also 
\textit{a skip-list proof for each of the queried blocks}. 
This results in higher communication cost required for an audit in DSCS I
--- which is expected and justified as DSCS I handles data dynamics.

\begin{table}[htbp]
\setlength{\tabcolsep}{1.2pt}
\centering
\caption{Storage overhead for the server and communication cost in DPDP I}{
\begin{tabular}{|c|c|c|c|c|c|c|c|c|}
\hline
File size & Storage cost for tags & Storage & \multirow{2}{*}{$|Q|$} & \multicolumn{4}{c|}{Communication cost (KB)} \\
\cline{5-8}
(MB) & and skip list (KB) & overhead &   & Audit & Insert  & Modify & Delete  \\
\hline\hline
1 & 0.71 & 0.07\% & 2 & 1.70   & 1.20 & 1.10 & 1.00 \\
\hline
10 & 7.37  & 0.07\% & 10 & 11.30  & 1.20 & 1.10 & 1.00\\
\hline
50 & 34.70 & 0.07\% & 10 & 12.50  & 1.20 & 1.10 & 1.00\\
\hline
200 & 141.90 & 0.07\% & 112 & 144.48 & 1.20 & 1.10 & 1.00\\
\hline
500 & 354.90 & 0.07\% & 112 & 146.98 & 1.20 & 1.10 & 1.00\\
\hline
\end{tabular}}
\label{tab:storage_commn_dpdpI}
\end{table}

\begin{table}[ht]
\setlength{\tabcolsep}{1.2pt}
\centering
\caption{Computation cost for the client $C$ and the server $S$ in DPDP I}
\begin{tabular}{|c|c|c|c|c|c|c|c|c|c|c|c|}
\hline
{File } & {Outsource} & \multirow{3}{*}{$|Q|$} & {Challenge} & {Prove} & {Verify} &
\multicolumn{2}{c|}{Insert }&
\multicolumn{2}{c|}{Delete }&
\multicolumn{2}{c|}{Modify }\\
{size} & {(sec)} & & {(msec)} & {(sec)} & {(sec)} &
\multicolumn{2}{c|}{(sec)}&
\multicolumn{2}{c|}{(sec)}&
\multicolumn{2}{c|}{(sec)}\\
\cline{2-2}
\cline{4-12}
(MB) & $C$ &   & $C$ & $S$ & $C$ & $C$ & $S$ & $C$ & $S$ & $C$ & $S$  \\
\hline
\hline
1 & 10.33 & 2 & 0.01 & 0.08 & 0.002 & 4.89 & 0.01 & 0.001 & 0.01 & 4.89 & 0.01 \\
\hline
10 & 83.099 & 10 & 0.02 & 0.37 & 0.004 & 4.99 & 0.09 & 0.001 & 0.08 & 4.99 & 0.09 \\
\hline
50 & 445.26 & 10 & 0.03 & 0.36 & 0.005  & 5.01 & 0.38 & 0.002 & 0.30 & 5.01 & 0.38 \\
\hline
200 & 1640.05 & 112 & 0.28 & 3.91 & 0.025 & 4.53 & 5.39 &  0.002 & 4.93 & 4.53  & 5.39 \\
\hline
500 & 3950.26 & 112 & 0.28 & 4.17 & 0.025 & 4.87 & 10.29 &  0.002 & 8.23 & 4.87 & 10.29  \\
\hline
\end{tabular}
\label{tab:comp_dpdpI}
\end{table}

\noindent
\textbf{Comparison between DSCS I and DPDP I}:\quad
We compare the performance of DSCS I with that of DPDP I~\cite{Erway_TISSEC} 
as both of them handle dynamic data, offer
public verifiability and are secure in the standard model.
Moreover, like~\cite{Erway_TISSEC}, DSCS I uses a rank-based authenticated skip list as the authenticated data structure.
We implement DPDP I and report the experimental results in Table~\ref{tab:storage_commn_dpdpI}
and Table~\ref{tab:comp_dpdpI}.
For ease of comparison, we keep the parameters same as DSCS I (Table~\ref{tab:storage_commn_dscsI} and Table~\ref{tab:comp_dscsI}).
From these tables, 
we observe that, although DPDP I performs slightly better than DSCS I, the overall performance of DSCS I is comparable with that of DPDP I.
However, we stress that, in this work, we look at the possibility and issues of constructing
DSCS protocols using secure network coding techniques.

\begin{table}[t]
\setlength{\tabcolsep}{1.2pt}
\centering
\caption{Storage overhead for the server $S$ and communication cost in DSCS II}{
\begin{tabular}{|c|c|c|c|c|c|}
\hline
File size  & Storage cost & Storage & \multirow{2}{*}{$|Q|$} & \multicolumn{2}{c|}{Communication cost (B)}\\
\cline{5-6}
(MB) & for tags (KB) & overhead &  & Audit & Append \\
\hline\hline
1 & 0.39 & 0.04\%  & 2 & 131 & 131 \\
\hline
10 & 2.751 & 0.03\%  & 10 & 131 & 131 \\
\hline
50 & 13.23 & 0.03\% & 10 & 131 & 131 \\
\hline
200 & 52.53 & 0.03\% & 112 & 131 & 131 \\
\hline
500 & 131.00 & 0.03\% & 112 & 131 & 131 \\
\hline
\end{tabular}}
\label{tab:storage_dsdcII}
\end{table}

\begin{table}[t]
\setlength{\tabcolsep}{1.2pt}
\centering
\caption{Computation cost for the client $C$ and the server $S$ in DSCS II}
\begin{tabular}{|c|c|c|c|c|c|c|c|}
\hline
{File } & {Outsource} & \multirow{3}{*}{$|Q|$} & {Challenge} & {Prove} & {Verify} &
\multicolumn{2}{c|}{Append }\\
{size} & {(sec)} & & {(msec)} & {(sec)} & {(sec)} &
\multicolumn{2}{c|}{(sec)}\\
\cline{2-2}
\cline{4-8}
(MB) & $C$ &   & $C$ & $S$ & $C$ & $C$ & $S$ \\
\hline
\hline
1 & 154.05 & 2 &  0.09 & 0.12 & 0.02  & 65.75 & 0.02 \\
\hline
10 & 1459.06 & 10 &  0.32 & 0.75 & 0.03  & 65.49 & 0.24 \\
\hline
50 & 7212.21 & 10 &  0.47 &  0.79 & 0.05  & 65.45 & 1.15 \\
\hline
200 & 28600.80 & 112 &  3.78 & 12.85 & 0.37 & 65.34 & 8.90 \\
\hline
500 & 96001.83 & 112 &  4.43  & 13.49 & 0.39 & 65.51 & 20.40 \\
\hline
\end{tabular}
\label{tab:comp_dscsII}
\end{table}

\subsection{Experimental Results for DSCS II} 
\label{sec:exp_DSCSII}

The additional storage cost (storage overhead at the server $S$) 
for DSCS II 
accounts for the authentication tags only.
The storage overhead is reported in Table~\ref{tab:storage_dsdcII}.
After the initial outsourcing of the data file, the client $C$ and the server $S$ need to communicate with each other either 
during an audit or during an append.
The communication cost during an audit for $Q$ data blocks includes 
the size of the aggregated tag (the size of the single aggregated block is not reported here). 
The communication costs for an audit and an append are reported in Table~\ref{tab:storage_dsdcII}.
Table~\ref{tab:comp_dscsII} summarizes the computation cost for DSCS II incurred during the initial outsourcing, an audit and an append. 
We note that the time needed to generate a tag is much more expensive (compared to DSCS I)
due to the costly operations in bilinear groups implemented using the PBC library.

\section{Conclusion}
\label{sec:conclusion}
In this work, we have proposed a secure cloud storage protocol for dynamic data (DSCS I) based on a secure network coding (SNC) protocol.
To the best of our knowledge, this is the first SNC-based DSCS protocol that is secure in the standard model and enjoys public verifiability.
We have discussed some challenges while constructing an efficient DSCS protocol from an SNC protocol.
We have also identified some limitations of an
SNC-based secure cloud storage protocol for dynamic data. However, some of these limitations follow from the underlying
SNC protocol used. A more efficient SNC protocol can give us a DSCS protocol with better efficiency.
We have also identified certain SNC protocols suitable for append-only data and constructed an efficient DSCS protocol (DSCS II) for append-only data.
We have shown that DSCS II overcomes some limitations of DSCS I.
Finally, we have provided prototype implementations of DSCS I and DSCS II in order to show their practicality
and compared the performance of DSCS I with that of an SNC-based secure cloud storage for static data
and that of DPDP I.


\section*{Appendix}
\appendix

\section{On Generic Construction of an Efficient DSCS Protocol from an SNC Protocol}
\label{app:chal_dscs}
We identify certain challenges towards providing a generic construction of an \textit{efficient} DSCS protocol
from an SNC protocol. 
We describe these challenges as follows.

\begin{enumerate}

\item \textit{The DSCS protocol must handle the varying values of $m$ appropriately}.
In a network coding protocol, the sender splits the file into $m$ vectors (or blocks) and
augments them with unit coefficient vectors before sending them into the network.
The length of these coefficient vectors is $m$ which remains constant during transmission.
On the other hand, in a DSCS protocol, the number of vectors may vary (for an insertion and a deletion).
Therefore, both the client and the server need to keep the latest value of $m$.

For a privately verifiable DSCS protocol,
the client and the server keep the up-to-date value of $m$.
For a publicly verifiable DSCS protocol,
the client includes the value of $m$ in her public key
and updates its value for each authenticated insertion and deletion. 
Thus, its latest value is known to the third party auditor (TPA) as well.

\item \textit{The index of a vector should not be embedded in its authentication tag}. In an SNC protocol,
the file to be transmitted is divided into $m$ vectors
$\hbox{v}_1,\hbox{v}_2,\ldots,\hbox{v}_m$, where each  $\hbox{v}_i\in{\mathbb{F}}^n$
for $i\in[1,m]$ (${\mathbb{F}}$ is replaced by ${\mathbb{Z}}$ in~\cite{Gennaro_PKC}).
The sender augments $\hbox{v}_i$ to form 
$\hbox{u}_i=[\hbox{v}_i~\hbox{e}_i]\in{\mathbb{F}}^{n+m}$ for $i\in[1,m]$, where
$\hbox{e}_i$ is the $m$-dimensional unit vector containing 1 in $i$-th
position and 0 in others. Let $V\subset{\mathbb{F}}^{n+m}$ be the linear subspace
spanned by these augmented basis vectors $\hbox{u}_1,\hbox{u}_2,\ldots,\hbox{u}_m$.
The sender authenticates the subspace $V$ by authenticating these augmented vectors before transmitting them to
the network. 
In a SNC protocol based on homomorphic MACs~\cite{HMAC_ACNS}, the sender generates a MAC for
the $i$-th basis vector $\hbox{u}_i$ and the index $i$ serves as an input to the pseudorandom function used to generate the MAC.
On the other hand, for some protocols based
on homomorphic signatures, $i$ is embedded in the tag $t_i$ for the $i$-th
augmented vector. 
For example, $H(\texttt{fid}||i)$ is embedded in $t_i$~\cite{BFKW_PKC,Gennaro_PKC},
where \texttt{fid} is the file identifier and $H$ is a hash function. 
Appendix~\ref{snc_exmpl} gives a brief overview 
of the protocol in~\cite{BFKW_PKC} and shows how $t_i$
includes the value $H(\texttt{fid}||i)$ (see Eqn.~\ref{eqn_Boneh}).

These SNC protocols are not suitable for the construction of an \textit{efficient} 
secure cloud storage protocol for generic dynamic data due to the following reason.
For generic dynamic data, the client can insert (or delete) a vector after (or from) a specified index (say, $i$) of the file. 
In both cases, the indices of the subsequent vectors (which were previously residing at positions $i+1,i+2,\ldots,m$) are changed. 
Therefore, in order to update the authentication tags for all these subsequent vectors,
the client has to download these vectors and compute the new tags before uploading the tags
to the cloud server. 
This makes the DSCS protocol inefficient. 
However, in a few SNC protocols, instead of embedding vector indices in tags, there is a one-to-one
mapping from the set of indices to some group~\cite{Charles,Dario_PKC}, and these
group elements are made public. 
Efficient DSCS protocols can be constructed from these SNC protocols. In fact, we construct
a DSCS protocol (described in Section~\ref{modified_scheme}) based on the SNC protocol
proposed by Catalano et al.~\cite{Dario_PKC}.

\item \textit{The freshness of data must be guaranteed}.
The freshness of storage requires that the server is storing an up-to-date version of the data.
For dynamic data,
the client can modify an existing vector. However, a malicious
server may discard the update and keep an old copy of the vector. 
As the old copy and its corresponding tag are valid, the client has no way to detect
if the server is storing the latest copy.

We ensure the freshness of the client's data, in our DSCS construction,
using a rank-based authenticated skip list built over the authentication tags for the vectors.
In other words, the authenticity of the vectors is maintained by their tags,
and the integrity of the tags is in turn maintained by the skip list.
When a vector is inserted (or modified),
its tag is also updated and sent to the server. The server updates the skip list accordingly.
While deleting a vector, the server simply removes the corresponding tag from the skip list.
Finally, the server sends to the client a proof that it has performed the required updates properly.
In Section~\ref{skip_list}, we discuss rank-based authenticated skip lists
that we use in our construction. 

\end{enumerate}

\noindent
In addition, it is often desired that a DSCS protocol
has the following property.
\begin{enumerate}
\setcounter{enumi}{3}

\item \textit{Public verifiability}\quad For a publicly verifiable DSCS protocol,
any third party auditor (TPA) with the knowledge of the client's public key
can perform audits.
In an SNC protocol based on
homomorphic MACs, some secret information (e.g., the secret key of 
the pseudorandom function used in~\cite{HMAC_ACNS}) is needed to verify the authenticity of a vector ---
which restricts a DSCS protocol constructed using such an SNC
protocol to be privately verifiable only.

\end{enumerate}

\section{A Homomorphic Signature Scheme for Network Coding Proposed by Boneh, Freeman, Katz and Waters}
\label{snc_exmpl}
Boneh et al.~\cite{BFKW_PKC} propose a homomorphic signature scheme for network coding that is secure
in the random oracle model~\cite{BR_RO} under the co-computational Diffie Hellman (co-CDH) assumption. 
Let $\mathcal{G}=(G_1,G_2,G_T,e,\psi)$ be a bilinear group tuple,
where $G_1,G_2$ and $G_T$ are multiplicative cyclic groups of prime order $p$, and the functions
$e: G_1\times G_2\rightarrow G_T$ (bilinear map) and $\psi:G_2\rightarrow G_1$ are efficiently
computable (see Section~\ref{sec:blmap}).
Then, the co-computational Diffie Hellman (co-CDH) problem in $(G_1,G_2)$ is to compute
$g^x\in G_1$ given $g,h$ and $h^x$, where $g\in G_1$ and $h,h^x\in G_2$ (for some $x\in{\mathbb{Z}}_p$).

We say that the co-CDH assumption holds in $(G_1,G_2)$ if,
for any probabilistic polynomial-time
adversary $\mathcal{A}(1^\lambda)$, the probability
\begin{eqnarray*}
\Pr_{\substack{g\xleftarrow{R}G_1\\h,z\xleftarrow{R}G_2}}[a\leftarrow\mathcal{A}(g,h,z=h^x):a=g^x]
\end{eqnarray*}
is negligible in $\lambda$, where the probability is taken over the internal
coin tosses of $\mathcal{A}$ and the random choices of $g,h$ and $z$.
We briefly describe the algorithms involved in this scheme~\cite{BFKW_PKC}.

\begin{itemize}
\item KeyGen$(1^\lambda,m,n)$: Let $\mathcal{G}=(G_1,G_2,G_T,e,\psi)$ be a bilinear group tuple,
where $G_1,G_2$ and $G_T$ are multiplicative cyclic groups of prime order $p>2^\lambda$, and the functions
$e: G_1\times G_2\rightarrow G_T$ (bilinear map) and $\psi:G_2\rightarrow G_1$ are efficiently
computable. Choose $g_1,\ldots,g_n\xleftarrow{R} G_1\backslash\{1\}$,
$h\xleftarrow{R} G_2\backslash\{1\}$ and $\alpha\xleftarrow{R}{\mathbb{F}}_p$. Take $z=h^\alpha$.
Let $H: {\mathbb{Z}}\times{\mathbb{Z}}\rightarrow G_1$ be a hash function considered to be a random oracle.
The public key is $pk=(\mathcal{G},H,g_1,\ldots,g_n,h,z)$, and the private key is $sk=\alpha$.

\item TagGen$(V,sk,m,n,\texttt{fid})$: Given the secret key $sk$,
a linear subspace $V\subset{\mathbb{F}}_p^{n+m}$ spanned by the augmented vectors
$\hbox{u}_1,\hbox{u}_2,\ldots,\hbox{u}_m$
and a random file identifier
\texttt{fid} $\in\{0,1\}^\lambda$, the sender outputs the signature
$$t_i=\left(\prod\limits_{j=1}^{m}{H(\texttt{fid}||j)^{u_{i(n+j)}}}\prod\limits_{j=1}^{n}{g_j^{u_{ij}}}\right)^{\alpha}$$
for the $i$-th basis vector $\hbox{u}_i=[u_{i1},u_{i2},\ldots,u_{i(n+m)}]\in{\mathbb{F}}_p^{n+m}$ for each $i\in[1,m]$.

\item Combine$(\{\hbox{y}_i,t_i,\nu_i\}_{1\le i\le l},pk,m,n,\texttt{fid})$: Given the public key $pk$,
the file identifier \texttt{fid} and $l$ tuples (each consisting of a vector $\hbox{y}_i\in{\mathbb{F}}_p^{n+m}$,
a coefficient $\nu_i\in{\mathbb{F}}_p$ and a signature $t_i$), an intermediate node outputs the signature
$t=\prod_{i=1}^{l}{t_i^{\nu_i}}$ for another vector
$\hbox{w}=\sum\limits_{i=1}^{l}{\nu_i\cdot\hbox{y}_i}\in{\mathbb{F}}_p^{n+m}$.

\item Verify$(\hbox{w},t,pk,m,n,\texttt{fid})$: Given the public key $pk$, the unique file identifier
\texttt{fid}, a signature $t$ and a vector $\hbox{w}=[w_1,w_2,\ldots,w_{n+m}]\in{\mathbb{F}}_p^{n+m}$,
an intermediate node or the receiver node checks whether
\begin{equation*}
e(t,h)\stackrel{?}=e\left(\prod\limits_{j=1}^{m}{H(\texttt{fid}||j)^{w_{n+j}}}\prod\limits_{j=1}^{n}{g_j^{w_j}},z\right).
\end{equation*}
If the equality holds, it outputs 1; it outputs 0, otherwise.

\end{itemize}

We recall that in a secure cloud storage protocol (using secure network coding), the client divides
the file $F$ associated with \texttt{fid} into $m$ vectors (or blocks) each of them having $n$ segments.
The $i$-th vector $\hbox{v}_i$ is of the form $[v_{i1},\ldots,v_{in}]\in{\mathbb{F}}^n,\forall i\in[1,m]$.
For each vector $\hbox{v}_i$, the client forms $\hbox{u}_i=[\hbox{v}_i~\hbox{e}_i]\in{\mathbb{F}}^{n+m}$
by augmenting the vector $\hbox{v}_i$ with the unit coefficient vector $\hbox{e}_i$. If we use
the current SNC protocol~\cite{BFKW_PKC}, the client runs
$\text{TagGen}(V,sk,m,n,\texttt{fid})$ to produce a signature (authentication tag)
\begin{equation}\label{eqn_Boneh}
\begin{aligned}
t_i & =\left(\prod\limits_{j=1}^{m}{H(\texttt{fid}||j)^{u_{i(n+j)}}}\prod\limits_{j=1}^{n}{g_j^{u_{ij}}}\right)^{\alpha}\\
    & =\left(H(\texttt{fid}||i)\prod\limits_{j=1}^{n}{g_j^{u_{ij}}}\right)^{\alpha}\\
    & =\left(H(\texttt{fid}||i)\prod\limits_{j=1}^{n}{g_j^{v_{ij}}}\right)^{\alpha},
\end{aligned}
\end{equation}
for each vector $\hbox{u}_i$ ($i\in[1,m]$). We observe that the vector index $i$ is embedded in
the tag corresponding to the $i$-th vector. Therefore, the scheme is not suitable
for construction of a secure cloud storage for dynamic (in generic sense) data
as mentioned in Appendix~\ref{app:chal_dscs}.

\section{Rank-Based Authenticated Skip List}
\label{app:skip_list}

\begin{figure}[t]
\small
\centering
\fbox{\includegraphics[width=.57\textwidth]{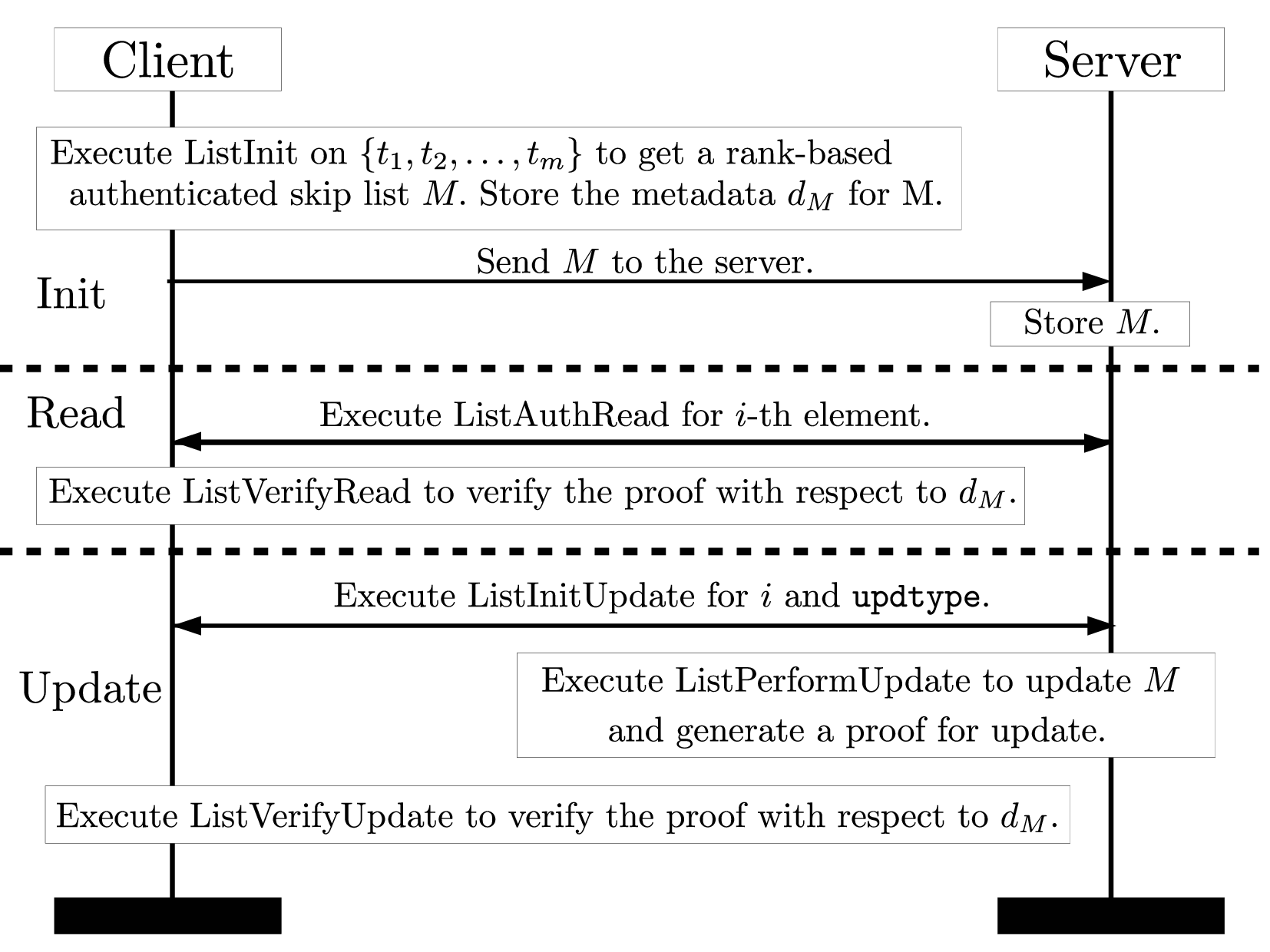}}
\caption{Workflow of a rank-based authenticated skip list built over 
	$\{t_1,t_2,\ldots,t_m\}$.}\label{fig:RBASL_workflow}
\end{figure}

We choose rank-based authenticated skip lists (over variants of Merkle hash trees)
to verify the freshness of data in our DSCS protocol due to the following reason.
In the dynamic versions of Merkle hash trees (e.g., authenticated red-black trees),
a series of insertions after a particular location
makes the tree imbalanced and increases the height of the tree by the number of such insertions.
In the two-party model, as in our case, no efficient rebalancing techniques
for such a tree have been studied~\cite{Erway_TISSEC}.
The algorithms for a rank-based authenticated skip list~\cite{Erway_TISSEC} stored remotely in a server
are as follows.
In Figure~\ref{fig:RBASL_workflow}, we show the overall workflow of a rank-based authenticated skip list $M$ 
built over $\{t_1,t_2,\ldots,t_m\}$, an ordered list of $m$ elements.

\begin{itemize}
\item
$\text{ListInit}(t_1,t_2,\ldots,t_m)$: Let $\{t_1,t_2,\ldots,t_m\}$ be an ordered list of $m$ elements 
on which a rank-based authenticated skip list $M$ is to be built. 
The construction of a rank-based authenticated skip list over $\{t_1,t_2,\ldots,t_9\}$
is shown in Figure~\ref{fig:RBASL_construction}(a)--(f).

\begin{figure*}[htbp]
\small
\centering
\fbox{\includegraphics[width=.48\textwidth,height=.27\textheight]{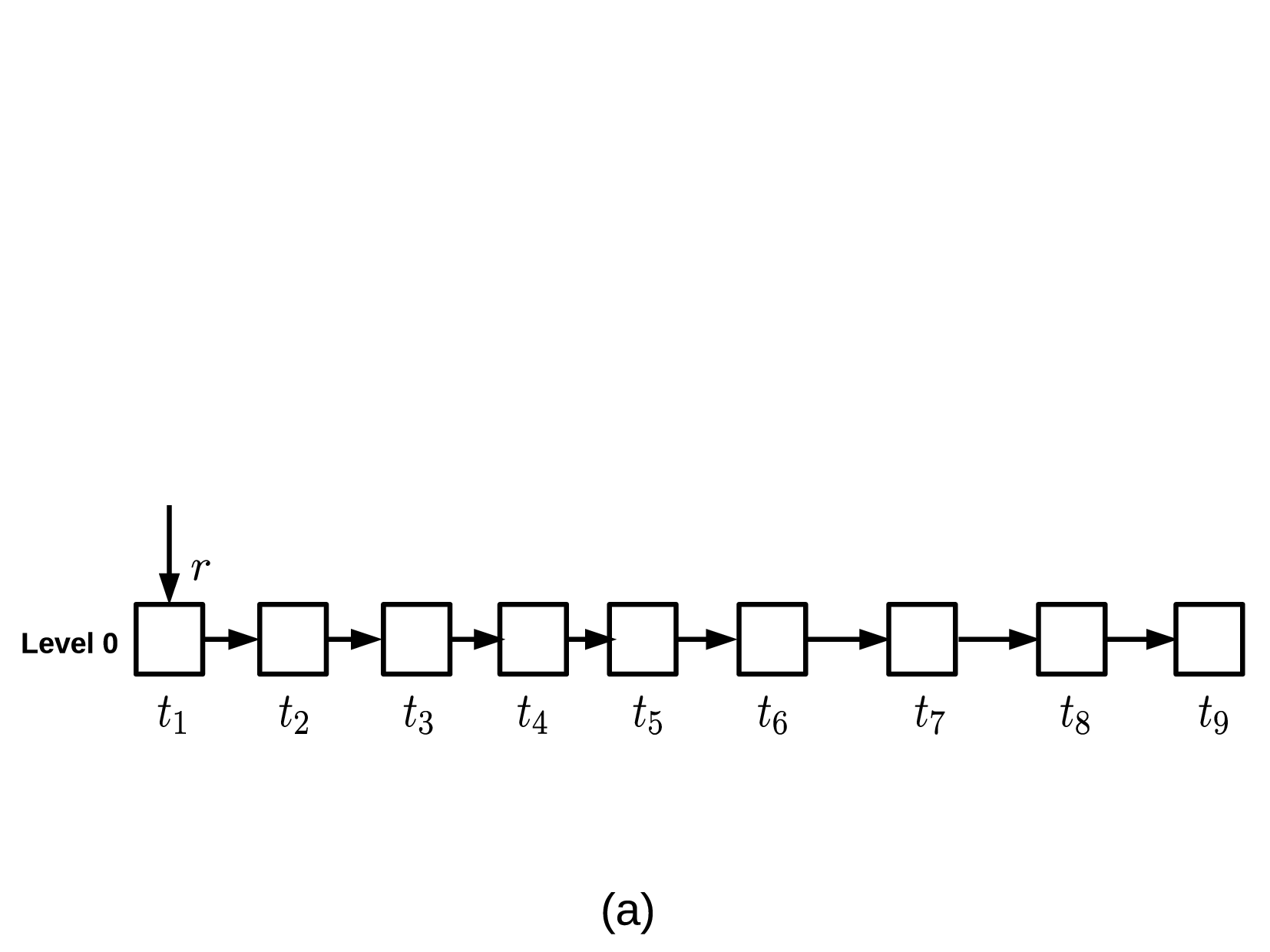}}
\fbox{\includegraphics[width=.48\textwidth,height=.27\textheight]{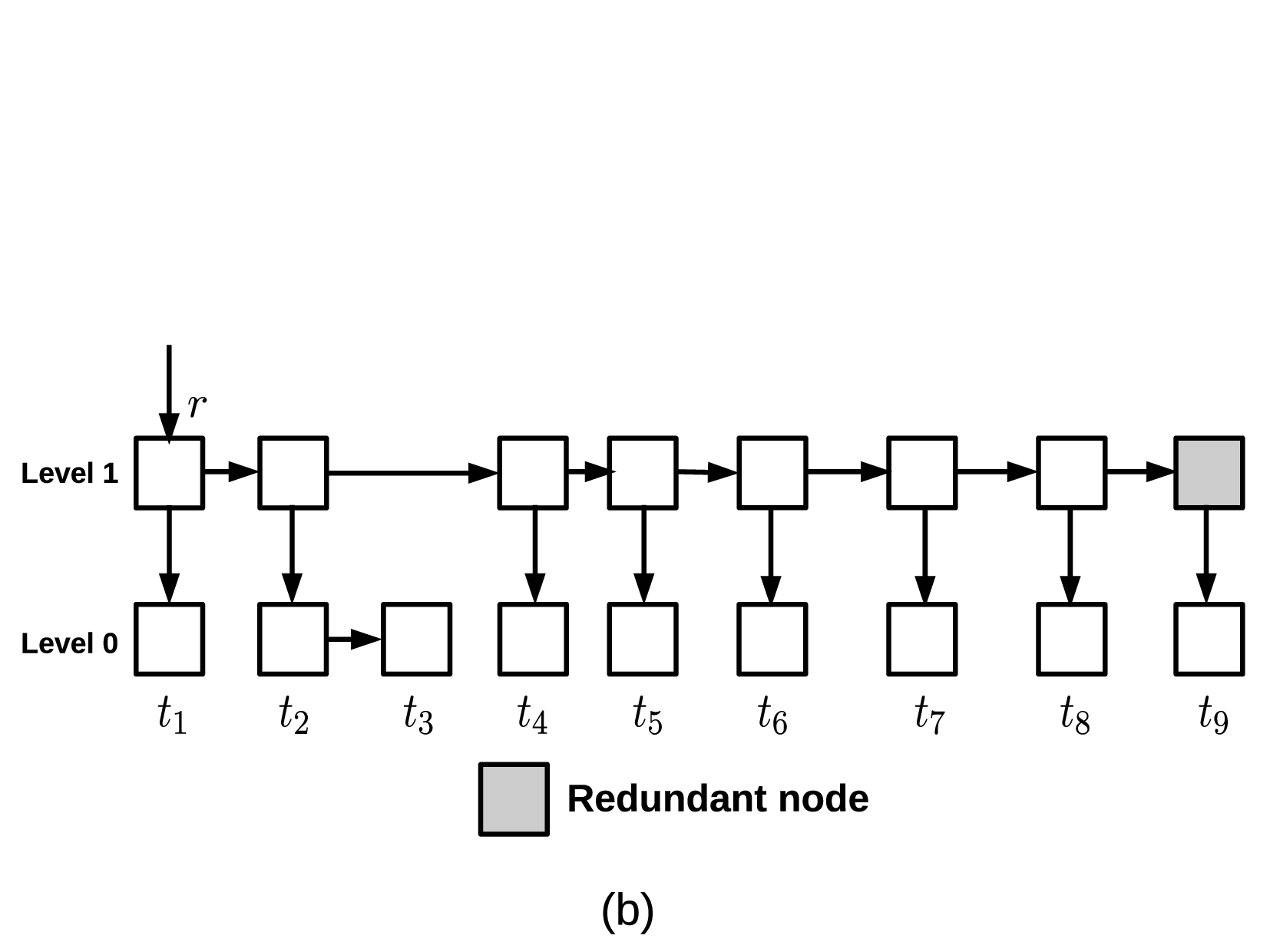}}
\fbox{\includegraphics[width=.48\textwidth,height=.27\textheight]{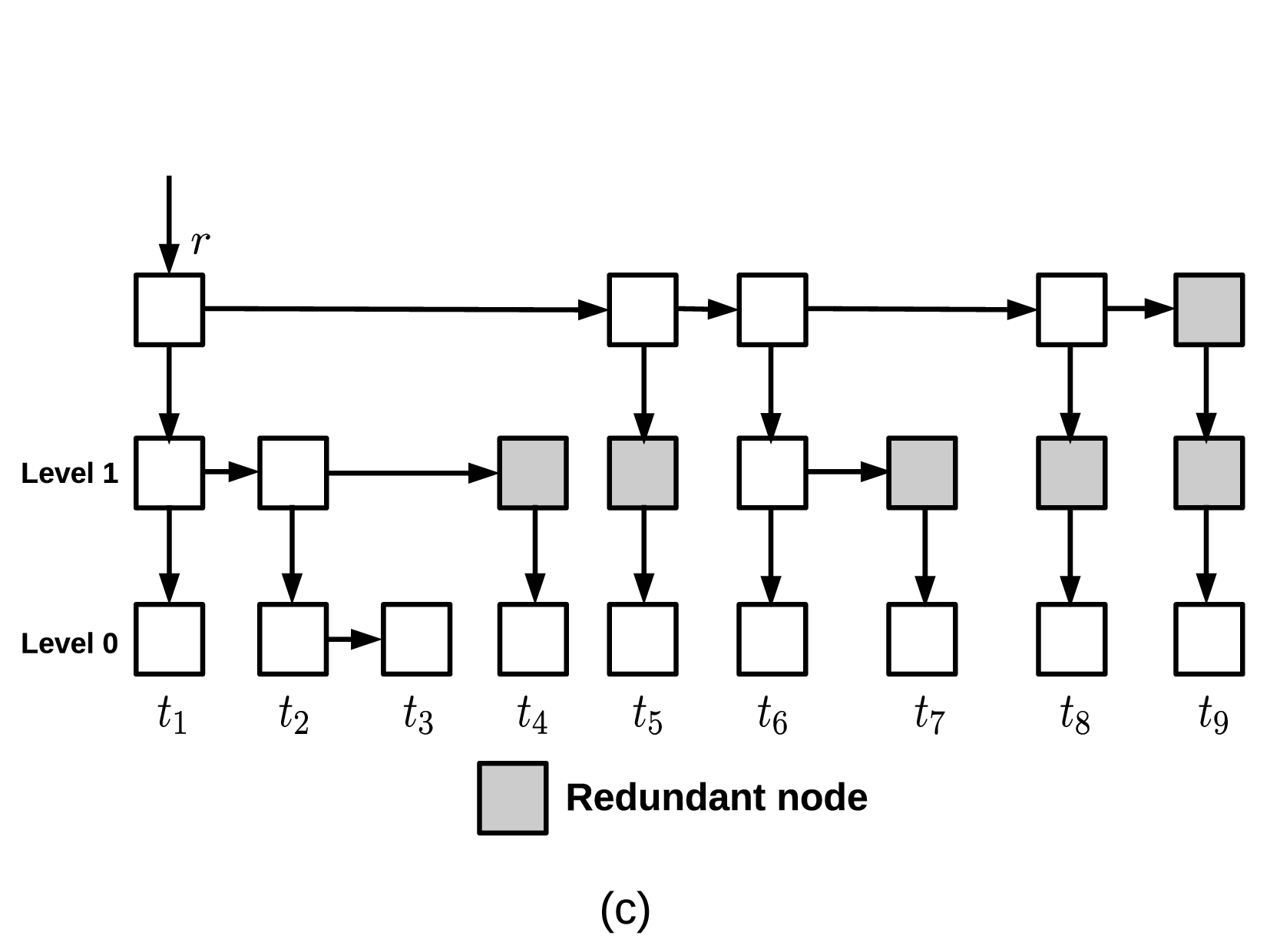}}
\fbox{\includegraphics[width=.48\textwidth,height=.27\textheight]{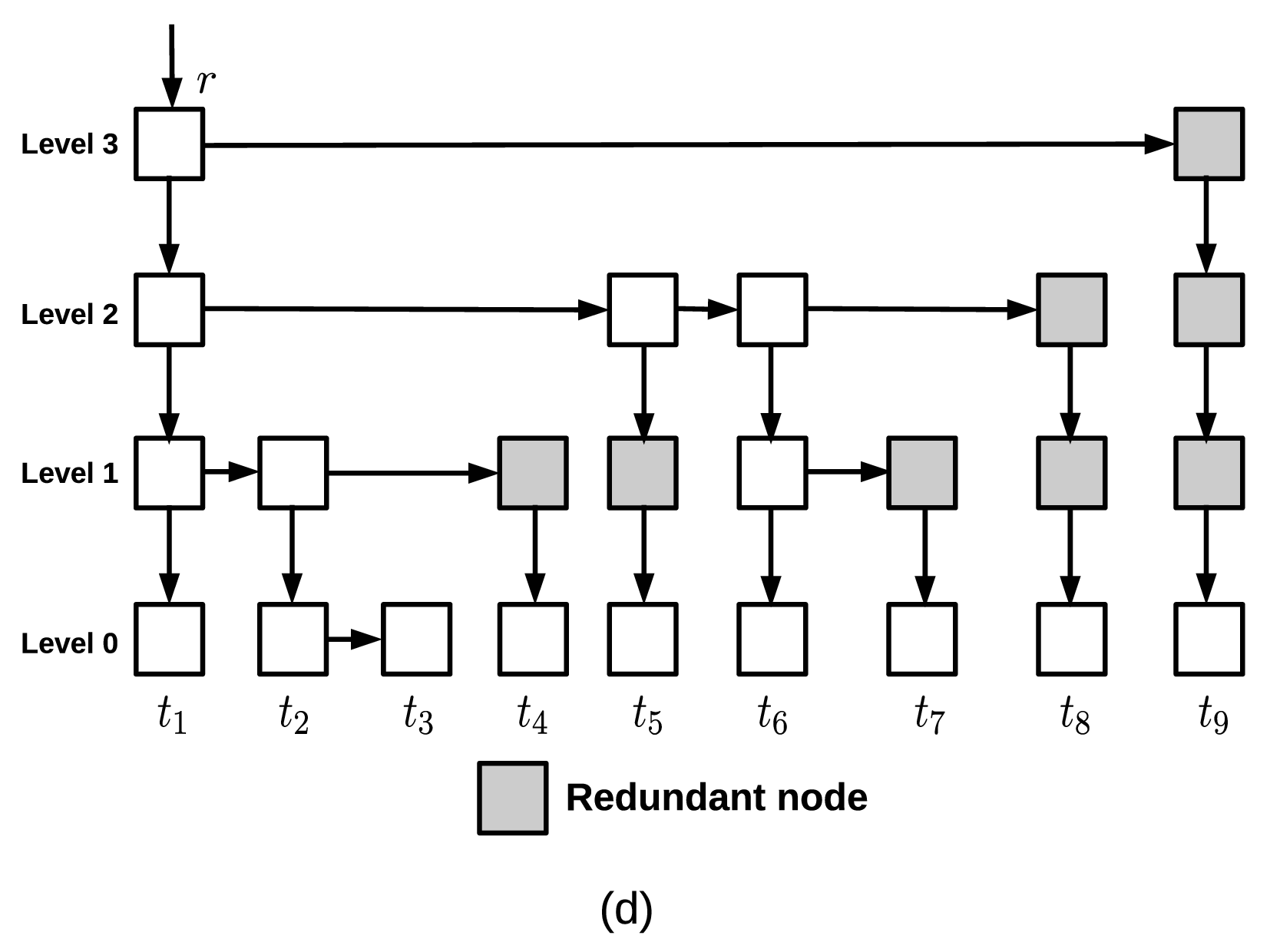}}
\fbox{\includegraphics[width=.48\textwidth,height=.27\textheight]{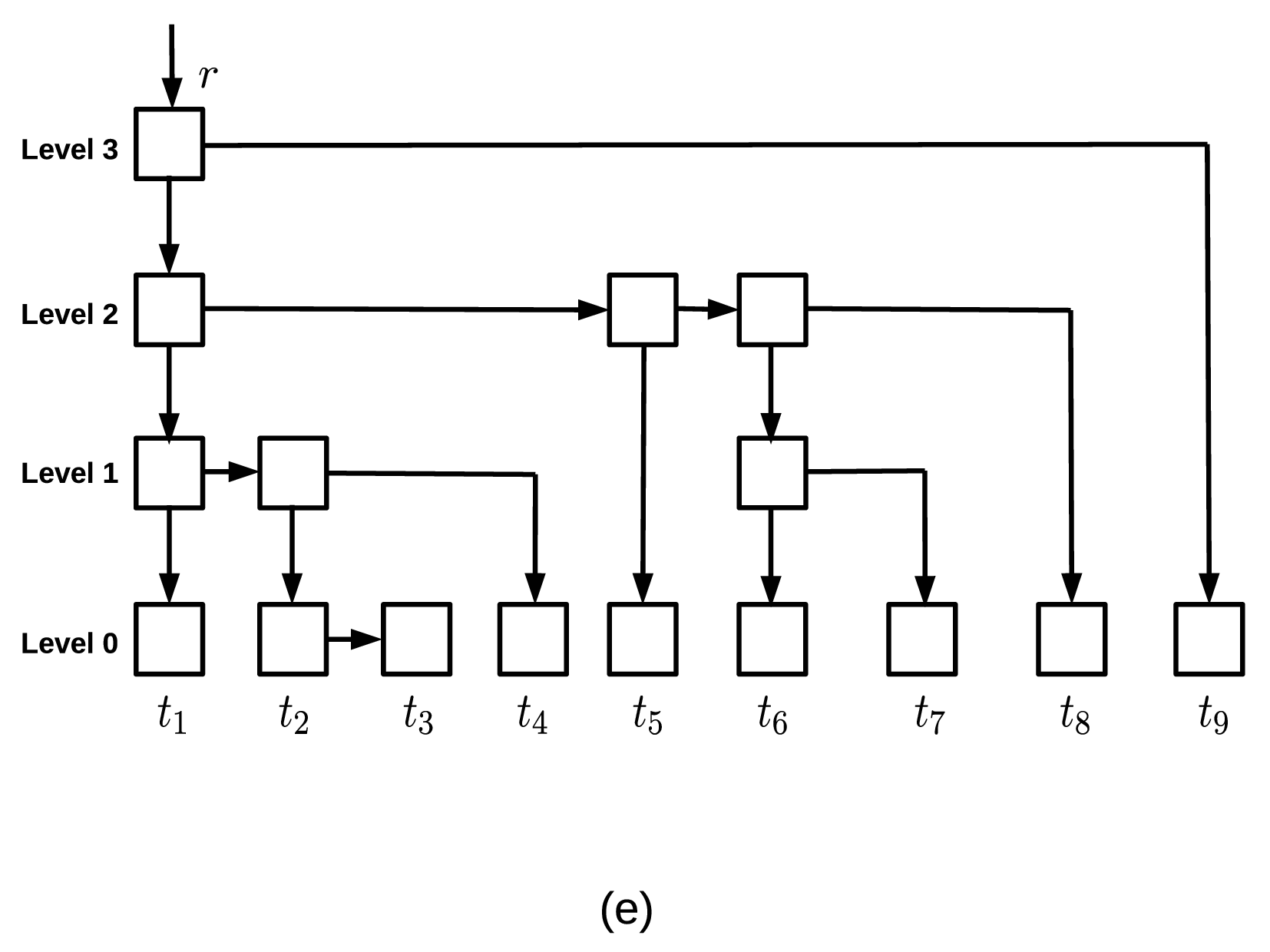}}
\fbox{\includegraphics[width=.48\textwidth,height=.27\textheight]{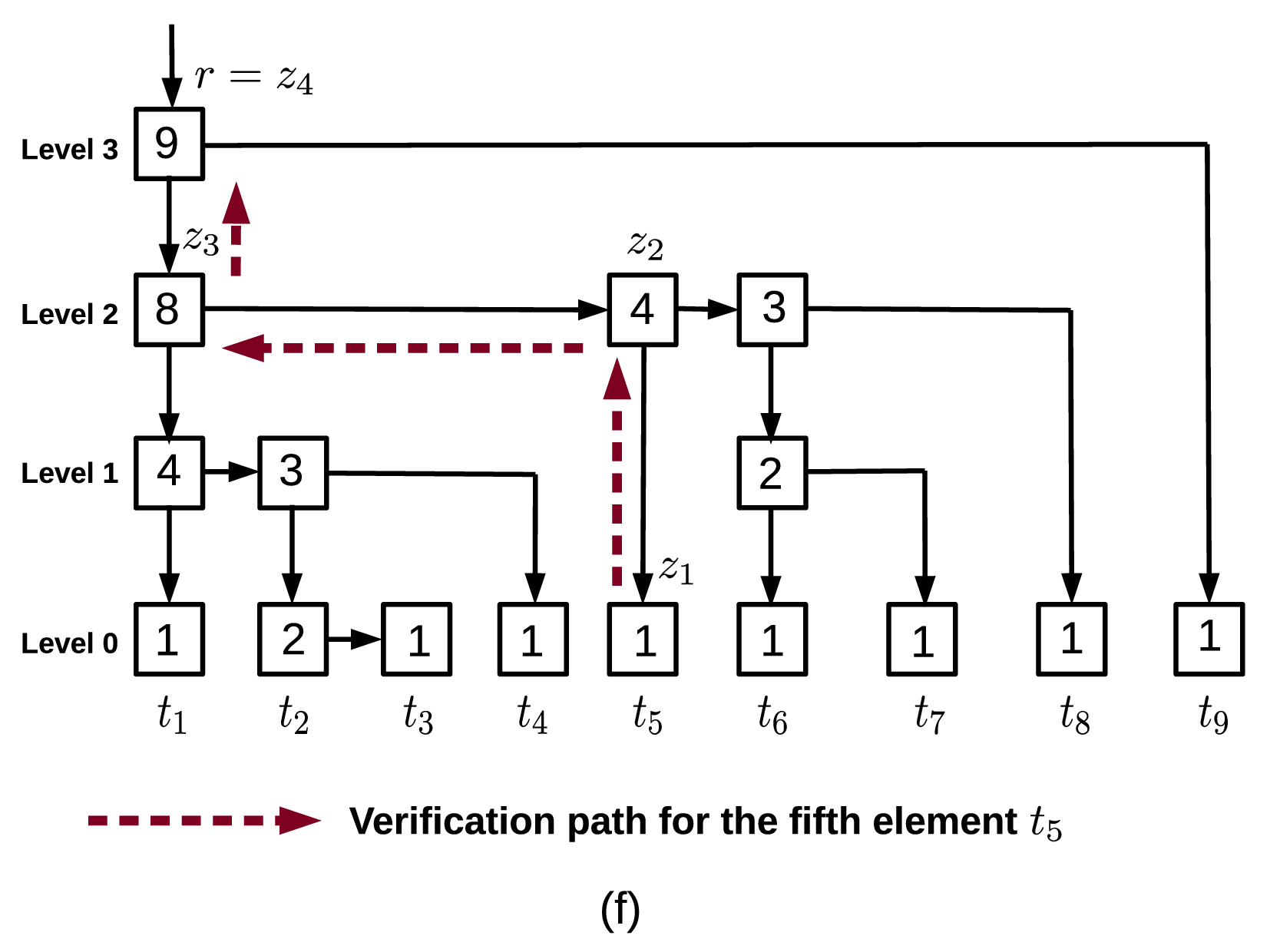}}
\caption{Construction of a rank-based authenticated skip list over 
	$\{t_1,t_2,\ldots,t_9\}$.
	The root node of the rank-based authenticated skip list is $r$.
	The $rank$ of each node in the final list is written inside it.
	The elements are in the bottom-level (Level 0) nodes,
	and the root node $r$ resides in the highest level (Level 3).
	The verification path for the fifth element $t_5$ is shown in the final list.}\label{fig:RBASL_construction}
\end{figure*}

Initially, the elements $\{t_1,t_2,\ldots,t_m\}$ are kept in a linked list as the bottom-level (Level 0) nodes 
(see Figure~\ref{fig:RBASL_construction}(a)).
Let $r$ denote the root node of the skip list.
For each node $z$ of the skip list: $right(z)$ and $down(z)$
are two pointers to the successors of $z$, $rank(z)$ is the number of bottom-level nodes reachable from $z$
(including $z$ if $z$ itself is a bottom-level node), $high(z)$
and $low(z)$ are the indices of the leftmost and rightmost bottom-level nodes reachable from
$z$, $f(z)$ is the label associated with the node $z$, and $l(z)$ is the level of $z$.
In addition, the $j$-th bottom-level node $z$ contains $x(z)=t_j,\forall j\in[1,m]$.
Initially, for each bottom-level node $z$, $l(z)$ is set to be 0. 
The down pointer $down(z)$ for each such node $z$
(except the rightmost node corresponding to $t_m$)
is set to be \texttt{null}.
For the rightmost node, both of its right and down pointers are set to be \texttt{null}.
Other fields of the nodes present in this initial skip list are set later.

Let us assume that the skip list is built up to $i$-th level where $i\ge 0$. 
We proceed to construct the $(i+1)$-th level of the skip list as follows.
We start with the leftmost node in the $i$-th level and move towards right until we reach the rightmost node
in the $i$-th level.
For each node $z$ in the $i$-th level, a bit $b$ is selected.
Let $z'$ be the predecessor node of $z$ in the $i$-th level (i.e., $right(z')=z$).
We note that $z'$ is \texttt{null} when $z$ is the leftmost node in the $i$-th level.
If the value of $b$ is 1, another node 
(say, $z''$ having $right(z'')$, $down(z'')$, $rank(z'')$, $high(z'')$, $low(z'')$, $f(z'')$  and $l(z'')$) 
corresponding to $z$ is added to the $(i+1)$-th level.
Let $z'''$ be the last node added to the $(i+1)$-th level before $z''$.
We note that $z'''$ is \texttt{null} when $z$ is the leftmost node in the $i$-th level.
The value of $l(z'')$ is set to be $(l(z)+1)$,
and three pointers are adjusted as follows: $down(z'')$ is set to be $z$, $right(z')$ is set to be \texttt{null},
and $right(z''')$ is set to be $z''$. 
The value of $b$ is taken to be 1 when $z$ is the leftmost or rightmost node in the $i$-th level
(i.e., a new node is added to the $(i+1)$-th level for each of them);
otherwise, the value of $b$ is chosen from $\{0,1\}$ uniformly at random.
We call a node ``redundant'' if it is not a bottom-level node and its right pointer is \texttt{null}.
We note that, while setting $right(z')$ to be \texttt{null}, the node $z'$ becomes
redundant (for $i\ge 1$). We remove each such redundant node $z'$ from the skip list
and let the right pointer of the predecessor node of $z'$ point to the successor of $z'$
present in the lower level (see Figure~\ref{fig:RBASL_construction}(d)--(e)).

For each level (starting from the bottom level), the following fields of each node $z$ in the skip list
are populated: $rank(z)$, $high(z)$ and $low(z)$.
Finally, for each node $z$, the label $f(z)$ is computed using a \textit{collision-resistant}
hash function $h$ as
\begin{equation}
f(z) = \begin{cases}
	   0  		& \text{if $z$ is }\texttt{null}\\
	   h(f_1)  	& \text{if }l(z)=0\\
	   h(f_2)  	& \text{if }l(z)>0,
       \end{cases}
\end{equation}
where $f_1=l(z)||rank(z)||x(z)||f(right(z))$ and $f_2=l(z)||rank(z)||f(down(z))||f(right(z))$.
Figure~\ref{fig:RBASL_construction}(f) illustrates a rank-based authenticated skip list
for an ordered list $\{t_1,t_2,\ldots,t_9\}$.

The rank-based authenticated skip list along with $m$ elements and all the associated information are stored in the server. 
The client stores only the value of $m$ and the label of the root node $r$ (i.e., $f(r)$) as the metadata $d_M$.

\item
$\text{ListAuthRead}(i,m,M)$: When the client sends a request to read the $i$-th element $t_i$, the server
sends the requested element along with a proof $\Pi(i)$ to the client. The server
computes the proof $\Pi(i)$ as follows.

We call the sequence of nodes $z_1,z_2,\ldots,z_k$ (where $z_1$ is the bottom-level node
storing the $i$-th element and $z_k=r$ is the root node of the skip list) the \textit{verification path}
for the $i$-th element in the skip list if
$z_1$ is reachable from the root node $z_k=r$ through the same nodes (but in the reverse order).
For example, the verification path for $t_5$ is the sequence of nodes $z_1,z_2,z_3,z_4$
as shown in Figure~\ref{fig:RBASL_construction}(f).
The proof $\Pi(i)$ computed by the server is of the form
\begin{equation}
\Pi(i)=(A(z_1),A(z_2),\ldots,A(z_k)),
\end{equation}
where $A(z)=(l(z),q(z),a(z),g(z))$.
Here, $l(z)$ is the level of the node $z$, $a(z)$ is 0 (or 1) if $down(z)$ (or $right(z)$)
points to the previous node of $z$ in the sequence, and $q(z)$ and $g(z)$ are the rank
and label (respectively) of the successor node of $z$ that is not present on the verification path.
We note that, given the proof $\Pi(i)$, one can compute the label of the root node $r$ (i.e., $f(r)$).

\item
$\text{ListVerifyRead}(i,d_M,t_i,\Pi(i),m)$: Upon receiving the proof $(t_i,\Pi(i))$ from the server,
the client checks if the proof corresponds to the latest metadata $d_M$ stored at her end.
The client outputs 1 if the proof matches with the metadata; she outputs 0, otherwise.

\item
$\text{ListInitUpdate}(i,\texttt{updtype},d_M,t_i',m)$: An update can be an insertion after
or a modification of or the deletion of the $i$-th bottom-level node. The type of the update is stored
in a variable \texttt{updtype}. The client defines $j=i$ (for an insertion or modification)
or $j=i-1$ (for a deletion). She executes $\text{ListAuthRead}(j,m,M)$ with the server (for the existing skip list $M$)
and verifies the response sent by the server by calling $\text{ListVerifyRead}(j,d_M,t_j,\Pi(j),m)$.
If the proof does not match with the metadata $d_M$ (the label of the root node of the existing skip list $M$),
she aborts.
Otherwise, she uses the proof $\Pi(j)$ to compute the metadata $d_M'$
that would be the new label of the root node if the server performs the update correctly
and updates the value of $m$ (if required).
The client stores $d_M'$ at her end temporarily.
Then, she asks
the server to perform the update specifying the location $i$, \texttt{updtype} (insertion,
deletion or modification) and the new element $t_i'$ (\texttt{null} for deletion).

\item
$\text{ListPerformUpdate}(i,\texttt{updtype},t_i',M)$: Depending on the value of \texttt{updtype},
the server performs the update asked by the client, computes a proof $\Pi$ similar
to the one generated during $\text{ListAuthRead}$ and sends $\Pi$ to the client.

\item
$\text{ListVerifyUpdate}(i,\texttt{updtype},t_i',d_M',\Pi,m)$: On receiving the proofs from
the server, the client verifies the proof $\Pi$ and computes the new metadata $d_{new}$ based on $\Pi$.
If $d_M'=d_{new}$ and $\Pi$ is a valid proof, the client sets $d_M=d_M'$, deletes the temporary value $d_M'$
and outputs 1. Otherwise, she changes $m$ to its previous value, deletes $d_M'$ and outputs 0.

\end{itemize}

\end{document}